Invariant graphical method for electron-atom scattering coupled-channel equations

J. B. Wang

School of Physics, The University of Western Australia, Australia

A. T. Stelbovics

Institute of Theoretical Physics, Curtin University of Technology, Australia

Abstract

In this paper, we present application examples of a graphical method for the efficient construction of potential matrix elements in quantum physics or quantum chemistry. The simplicity and power of this method are illustrated through several examples. In particular, a complete set of potential matrix elements for electron-Lithium scattering are derived for the first time using this method, which removes the frozen core approximation adopted by previous studies. This method can be readily adapted to study other many-body quantum systems.

# 1. Introduction

The fundamental problem encountered by a theoretician working in the field of quantum physics or quantum chemistry is to evaluate the so-called observables $\langle\psi|\Omega|\psi\rangle$ to certain precision. Here $\psi$ is the wave function of the system under investigation and $\Omega$ is the operator representing the physical quantity one needs to evaluate. This seemingly easy task encounters many obstacles, one of which is the complexity of the physical structure of a composite system. For example, there are several electrons and nuclei interacting with each other in an atomic or molecular system. Even though the exact form of such interactions is well known, it is exceedingly difficult to obtain accurate solutions from the corresponding Schrödinger or Dirac equation.

However, most physical systems have symmetry properties that influence their dynamics and structures in a way independent of the detailed interactions. A good example is the motion of a particle in a central field. The symmetry of the interaction leads to the conservation of angular momentum that is independent of the detailed form of the interaction potential V(r). Angular momentum theory deals with rotational symmetries of a composite system. In classical physics, angular momentum treatments are straightforward following the simple manipulation rules of vectors. In the quantum world, not only the magnitude of angular momenta is quantized, but also their spatial orientations.

Angular momentum calculations in quantum theory prove to be most laborious, especially when many-body systems or complicated operators are involved. Furthermore, the final expressions are often incomprehensible even for experienced researchers in the field (Williams *et al.* 1992, Wang *et al.* 1995). Unless one derives the equations himself by a painstakingly tedious procedure, one can not really tell where the phases and the n-j symbols come from. It is also quickly apparent once one begins to derive the complete reduced potential matrix-elements for systems beyond four electrons that the expressions become rapidly intractable to algebraic derivation by conventional methods. The problems are simply due to the vast number of Clebsch-Gordan coefficients that comprise the expressions, each of which is summed over several angular momentum labels. In addition, there are many pages of practical formulas that one often needs in order to perform these calculations, such as those listed in the appendixes of Brink and Satchler (1993).

In view of these difficulties, we started to search for a method that is transparent, free of the incomprehensible phases, makes minimal use of formulas, and leads easily to a correct final expression. We will demonstrate in this paper that the invariant graphical method initiated by Danos and Fano (1998) meets all the above criteria. We will then apply this method to derive the reduced matrix elements of selected electron-atom interaction potentials, which are not available mainly due to the difficulties mentioned above. This method can be readily adapted to study other many-body systems such as two-photon double ionization of atoms (Kheifets 2007, 2009), two-nucleon knockout reactions (Simpson *et al.* 2009), spin networks (Aquilanti *et al.* 2009), and entangled N spin-1/2 qubits (Suzuki *et al.* 2009; Maser *et al.* 2009).

A central point to this method is the concept that most physical quantities are invariant, i.e. they are independent of the coordinate system one chooses for their description. This is the case for energies, cross-sections, transition probabilities. In other words, these quantities are scalar quantities and have zero angular momentum. Consequently, the tensors they may contain would couple to a total of zero angular momentum. We will demonstrate, in the next section, that angular momentum calculations become much simpler once we take this invariance into account. Atomic units are used throughout this paper.

## 2. Elements of the invariant graphical method

In the invariant graphical method, there is only one basic graph (Fig. 1a) for angular momentum coupling and recoupling. The essential elements are the horizontal lines, each representing a tensor with a given angular momentum labeled above the line. Here tensors are used to represent all quantities including wave functions, operators, amplitudes, etc. In this way, a scalar quantity is a tensor of rank zero. A vector is a tensor of rank one and so on. As an example, the state with angular momentum L can be represented by a tensor $\psi$ of the rank L. This graph provides the basic transformation associated with four angular momenta to re-arrange them into a different coupling scheme. The transformation is represented by a recoupling box, mathematically associated with a square 9-j symbol such that

$$\left[\left[A^{[a]}B^{[b]}\right]^{[e]}\left[C^{[c]}D^{[d]}\right]^{[f]}\right]^{[i]} = \sum_{gh} \begin{vmatrix} a & b & e \\ c & d & f \\ g & h & i \end{vmatrix} \left[\left[A^{[a]}C^{[c]}\right]^{[g]}\left[B^{[b]}D^{[d]}\right]^{[h]}\right]^{[i]}, \qquad (1)$$

where $\left[A^{[J]}B^{[K]}\right]^{[I]}_{M} = \sum_{m_J m_K}(JKm_Jm_K|IM)A^{[J]}_{m_J}B^{[K]}_{m_K}$ is represented by [diagram with J, I, K labels]. The square 9-j symbol is related to the Wigner 9-j coefficients by

$$\begin{vmatrix} a & b & e \\ c & d & f \\ g & h & i \end{vmatrix} = \hat{e}\hat{f}\hat{g}\hat{h} \begin{Bmatrix} a & b & e \\ c & d & f \\ g & h & i \end{Bmatrix}, \qquad (2)$$

where $\hat{a} = \sqrt{2a+1}$.

This basic graph is the key building block of this method. All possible recoupling transformation can be performed based upon this graph. For example, the recoupling of three angular momenta can be derived using the graph illustrated in Fig. 1b as

$$\left[B^{[b]}\left[C^{[c]}D^{[d]}\right]^{[f]}\right]^{[i]} = \sum_h \begin{vmatrix} 0 & b & b \\ c & d & f \\ c & h & i \end{vmatrix} \left[C^{[c]}\left[B^{[b]}D^{[d]}\right]^{[h]}\right]^{[i]}. \tag{3}$$

In this method, there is no need to use 6-j symbols, which carry extra phases and normalization constants that can lead to tedious book-keeping and are prone to errors. There is also no need to remember the many sum rules, which are required to eliminate unnecessary summation indexes in other conventional methods.

## 3. Applications

Coupled channel equations have been used extensively in the description of electron-atom scattering. Their application has been particularly successful in the so-called convergent close-coupling approach (CCC) implemented by Bray and Stelbovics (1992) and subsequently extended to other atoms with one or two valence electrons outside an inert core (Bray *et al*. 1999, Fursa and Bray 1995, 2003). To perform such calculations, one requires the reduced matrix elements of the coupled channel potentials, which we denote by $\langle \mathsf{L'} \| V \| \mathsf{L} \rangle$ with $\mathsf{L}$ representing the complete set of quantum numbers specifying a particular configuration of the system under study. For electron scattering from an *N*-electron target, the coupled channel reduced matrix elements have the following structure

$$\langle \mathsf{L'} \| V \| \mathsf{L} \rangle = \left\langle \mathsf{L'} \left\| -\frac{N}{r_0} + \sum_{i=1}^{N}\frac{1}{r_{0i}} - \left(-\sum_{i=0}^{N}\frac{N}{r_i} + \sum_{\substack{i,j=0 \\ i>j}}^{N}\frac{1}{r_{ij}} - E\right)\sum_{i=1}^{N}P_{0i} \right\| \mathsf{L} \right\rangle,$$

which includes both the direct and exchange terms. Here *P* is the permutation operator, E is the system energy, subscript *0* denotes the incoming electron, and subscripts *1-N* denote the target electrons. Using the explicitly antisymmetrised target states, the reduced matrix elements can be simplified to

$$\langle \mathsf{L'} \| V \| \mathsf{L} \rangle$$
$$= \left\langle \mathsf{L'} \left\| N\left(-\frac{1}{r_0} + \frac{1}{r_{01}}\right) - \left(\frac{1}{r_{01}} + \frac{N-1}{r_{02}} + \frac{N-1}{r_{12}} + \frac{(N-1)(N-2)}{2r_{23}} - N\left(E + \frac{N}{r_0}\right)\right)P_{01} \right\| \mathsf{L} \right\rangle.$$

Each of these terms will be derived explicitly for e-He and e-Li scattering in sections (3.2) and (3.3).

By employing the invariant graphical method, we will derive a complete set of direct and exchange potential matrix elements for electron-atom scattering, which are beyond the frozen core approximation adopted by Wu *et al.* (2004), Fursa *et al.* (2003), Bray *et al.* (1999, 1994) and Zhang *et al.* (1992).

### 3.1 Two-electron system

To illustrate the use of the invariant graphical method, we first consider the direct Coulomb matrix elements for a two-electron system, i.e. $\langle (\ell'_a \ell'_b)\ell \| 1/r_{12} \| (\ell_a \ell_b)\ell \rangle$, which was also discussed in detail in Danos and Fano (1998). The complete coupling and recoupling graph is shown in Fig. 2. As described earlier, each line in the figure represents a tensor with a given angular momentum indicated by the label above the line. The join of two lines into one line represents coupling of their angular momenta. Recoupling into a different tensor set is indicated by a transformation box and needs to be carried out until each component of the graph is expressible in terms of single-particle matrix elements (e.g. the end boxes $[\lambda |\ell'_a| \ell_a]$ and $[\lambda |\ell'_b| \ell_b]$ in this case). The invariance of such a triple product provides a means of eliminating unnecessary intermediate indexes. For example, it demands that the coupling of $\ell'_a$ and $\ell_a$ must give rise to a tensor with angular momentum $\lambda$ instead of an arbitrary index. It also helps in identifying the most direct and economical recoupling scheme. We first expand the two-body Coulomb potential into its multipole components

$$\frac{1}{r_{12}} = \sum_\lambda R_\lambda(r_1 r_2) \left[ Y^{[\lambda]} Y^{[\lambda]} \right]^{[0]}, \tag{4}$$

where $R_\lambda(r_1 r_2) = \frac{4\pi}{\hat{\lambda}^2} \frac{r_<^\lambda}{r_>^{\lambda+1}}$. The corresponding final expression can then be directly read off from Fig. 2 as

$$\begin{aligned}
&\left\langle (\ell'_a \ell'_b)\ell \left\| \frac{1}{r_{12}} \right\| (\ell_a \ell_b)\ell \right\rangle \\
&= \sum_\lambda R_\lambda(r_1 r_2) \, i^{\ell'_a + \ell'_b - \ell_a - \ell_b} \left\langle \left[ Y^{[\ell'_a]} Y^{[\ell'_b]} \right]^{[\ell]} \left\| \left[ Y^{[\lambda]} Y^{[\lambda]} \right]^{[0]} \right\| \left[ Y^{[\ell_a]} Y^{[\ell_b]} \right]^{[\ell]} \right\rangle \\
&= \sum_\lambda R_\lambda(r_1 r_2) i^{\ell'_a + \ell'_b - \ell_a - \ell_b} \frac{1}{\hat{\ell}} \begin{bmatrix} \ell'_a & \ell'_b & \ell \\ \ell_a & \ell_b & \ell \\ \lambda & \lambda & 0 \end{bmatrix} \begin{bmatrix} \lambda & \lambda & 0 \\ \lambda & \lambda & 0 \\ 0 & 0 & 0 \end{bmatrix} [\lambda|\ell'_a|\ell_a][\lambda|\ell'_b|\ell_b]
\end{aligned} \tag{5}$$

where $Y^{[L]} = (-i)^L Y_L$ and $Y_L$ is the regular spherical harmonic tensor. This definition is introduced to avoid unnecessary phases entering the coupling and recoupling scheme.

Each square 9-j symbol in Eq. (5) corresponds to a recoupling box in Fig. 2. The end-box represents an invariant matrix element containing three spherical harmonic tensors written as $[\ell|k|j]$, which equals

$$[\ell|k|j] = i^{\ell+k+j} \frac{\hat{\ell}\hat{k}\hat{j}}{\sqrt{4\pi}} \begin{pmatrix} \ell & k & j \\ 0 & 0 & 0 \end{pmatrix}. \tag{6}$$

For numerical computation purposes, Eq. (5) can be treated as a final expression. Nevertheless, further simplification can be made considering the many zeros in the 9-j symbols. This leads to the well-known result (see, for example, Cowan 1981)

$$\left\langle (\ell'_a \ell'_b)\ell \left\| \frac{1}{r_{12}} \right\| (\ell_a \ell_b)\ell \right\rangle$$

$$= \sum_\lambda R_\lambda(r_1 r_2)(-1)^{\lambda+\ell} \hat{\ell}'_a \hat{\ell}'_b \hat{\ell}_a \hat{\ell}_b \begin{pmatrix} \ell'_a & \lambda & \ell_a \\ 0 & 0 & 0 \end{pmatrix} \begin{pmatrix} \ell'_b & \lambda & \ell_b \\ 0 & 0 & 0 \end{pmatrix} \begin{Bmatrix} \ell'_a & \ell'_b & \ell \\ \ell_b & \ell_a & \lambda \end{Bmatrix}. \tag{7}$$

### 3.2 Electron-He scattering (a three-electron system)

As a second more complicated example, we will derive the momentum-space direct and exchange matrix elements for a three-electron system, such as electron scattering from a helium atom. The general configuration of the system can be represented by $|\mathsf{L}\rangle \equiv |k(\ell_0(\ell_1 \ell_2)\ell)L;(\sigma_0(\sigma_1 \sigma_2)s)\mathsf{S}\rangle$, where $k$, $\ell_0$ and $\sigma_0$ are respectively the linear, orbital angular and spin momentum of the incoming electron, $\ell_1$ and $\ell_2$ are the angular momenta of the target electrons, $\sigma_1$ and $\sigma_2$ are the spin momenta of the target electrons, $\ell$ and $s$ are the total orbit angular and spin momentum of the target atom, and $L$ and $\mathsf{S}$ are the total orbit angular and spin momentum of the entire system. Note that the linear radial integral part of the direct and exchange elements is straightforward to work out and has the general form of

$$\langle n'_2 \ell'_2, n_2 \ell_2 \rangle \langle k' \ell'_0, k\ell_0; n'_1 \ell'_1, n_1 \ell_1 \rangle$$
$$\equiv \left\langle \varphi_{n'_2 \ell'_2} \middle| \varphi_{n_2 \ell_2} \right\rangle \int u_{\ell'_0}(k' r_0) u_{\ell_0}(k r_0) \varphi_{n'_1 \ell'_1}(r_1) \varphi_{n_1 \ell_1}(r_1) \frac{r_<^\lambda}{r_>^{\lambda+1}} dr_0 dr_1 ,$$

where $\varphi_{n\ell}$ are the target atomic orbitals with quantum numbers $n$ and $\ell$, and $u_\ell$ is the $\ell$th partial wave of the projectile electron. The coupling and recoupling graph for the direct and exchange potential spin part is shown in Fig. 3 and those for the orbital part are shown in Figs. 4-8. The corresponding final expressions can then be directly read off from the figures as

$$\left\langle k'\big(\ell'_0(\ell'_1\ell'_2)\ell'\big)L;(\sigma'_0(\sigma'_1\sigma'_2)s')\mathsf{S}\left\|\frac{1}{r_{01}}\right\|k\big(\ell_0(\ell_1\ell_2)\ell\big)L;(\sigma_0(\sigma_1\sigma_2)s)\mathsf{S}\right\rangle$$

$$=\sum_\lambda \langle n'_2\ell'_2,n_2\ell_2\rangle\langle k'\ell'_0,k\ell_0;n'_1\ell'_1,n_1\ell_1\rangle\langle\sigma'_0(\sigma'_1\sigma'_2)|\sigma_0(\sigma_1\sigma_2)\rangle i^{\ell'_0+\ell'_1+\ell'_2-\ell_0-\ell_1-\ell_2}\times$$

$$\left\langle \left[Y^{[\ell'_0]}(\vec{r}_0)\left[Y^{[\ell'_1]}(\vec{r}_1)Y^{[\ell'_2]}(\vec{r}_2)\right]^{[\ell']}\right]^{[L]}\left\|\left[Y^{[\lambda]}(\vec{r}_0)Y^{[\lambda]}(\vec{r}_1)\right]^{[0]}\right\|\left[Y^{[\ell_0]}(\vec{r}_0)\left[Y^{[\ell_1]}(\vec{r}_1)Y^{[\ell_2]}(\vec{r}_2)\right]^{[\ell]}\right]^{[L]}\right\rangle$$

$$=\sum_\lambda \langle n'_2\ell'_2,n_2\ell_2\rangle\langle k'\ell'_0,k\ell_0;n'_1\ell'_1,n_1\ell_1\rangle i^{\ell'_0+\ell'_1+\ell'_2-\ell_0-\ell_1-\ell_2}\frac{\hat{\ell}_2}{\hat{L}}\delta_{\ell'_2\ell_2}[\lambda|\ell'_0|\ell_0][\lambda|\ell'_1|\ell_1]\times$$

$$\begin{bmatrix}\ell'_0 & \ell' & L \\ \ell_0 & \ell & L \\ \lambda & \lambda & 0\end{bmatrix}\begin{bmatrix}\ell'_1 & \ell'_2 & \ell' \\ \ell_1 & \ell_2 & \ell \\ \lambda & 0 & \lambda\end{bmatrix}\begin{bmatrix}\lambda & \lambda & 0 \\ \lambda & \lambda & 0 \\ 0 & 0 & 0\end{bmatrix},$$

(8)

$$\left\langle k'\big(\ell'_0(\ell'_1\ell'_2)\ell'\big)L;(\sigma'_0(\sigma'_1\sigma'_2)s')\mathsf{S}\left\|\frac{1}{r_{01}}P_{01}\right\|k\big(\ell_0(\ell_1\ell_2)\ell\big)L;(\sigma_0(\sigma_1\sigma_2)s)\mathsf{S}\right\rangle$$

$$=\sum_\lambda \langle n'_2\ell'_2,n_2\ell_2\rangle\langle k'\ell'_0,n_1\ell_1;n'_1\ell'_1,k\ell_0\rangle\langle\sigma'_0(\sigma'_1\sigma'_2)|P_{01}|\sigma_0(\sigma_1\sigma_2)\rangle i^{\ell'_0+\ell'_1+\ell'_2-\ell_0-\ell_1-\ell_2}\times$$

$$\left\langle\left[Y^{[\ell'_0]}(\vec{r}_0)\left[Y^{[\ell'_1]}(\vec{r}_1)Y^{[\ell'_2]}(\vec{r}_2)\right]^{[\ell']}\right]^{[L]}\left\|\left[Y^{[\lambda]}(\vec{r}_0)Y^{[\lambda]}(\vec{r}_1)\right]^{[0]}\right\|\left[Y^{[\ell_0]}(\vec{r}_1)\left[Y^{[\ell_1]}(\vec{r}_0)Y^{[\ell_2]}(\vec{r}_2)\right]^{[\ell]}\right]^{[L]}\right\rangle$$

$$=\sum_{\lambda q}\langle n'_2\ell'_2,n_2\ell_2\rangle\langle k'\ell'_0,n_1\ell_1;n'_1\ell'_1,k\ell_0\rangle i^{\ell'_0+\ell'_1+\ell'_2-\ell_0-\ell_1-\ell_2}\times$$

$$\frac{\hat{\ell}_2}{\hat{L}}\delta_{\ell'_2\ell_2}[\lambda|\ell'_0|\ell_1][\lambda|\ell'_1|\ell_0]\begin{bmatrix}0 & \tfrac{1}{2} & \tfrac{1}{2} \\ \tfrac{1}{2} & \tfrac{1}{2} & s \\ \tfrac{1}{2} & s' & \mathsf{S}\end{bmatrix}\begin{bmatrix}0 & \ell_0 & \ell_0 \\ \ell_1 & \ell_2 & \ell \\ \ell_1 & q & L\end{bmatrix}\begin{bmatrix}\ell'_0 & \ell' & L \\ \ell_1 & q & L \\ \lambda & \lambda & 0\end{bmatrix}\begin{bmatrix}\ell'_1 & \ell'_2 & \ell' \\ \ell_0 & \ell_2 & q \\ \lambda & 0 & \lambda\end{bmatrix}\begin{bmatrix}\lambda & \lambda & 0 \\ \lambda & \lambda & 0 \\ 0 & 0 & 0\end{bmatrix},$$

(9)

$$\left\langle k'\big(\ell'_0(\ell'_1\ell'_2)\ell'\big)L;(\sigma'_0(\sigma'_1\sigma'_2)s')\mathsf{S}\left\|\frac{1}{r_{02}}P_{01}\right\|k\big(\ell_0(\ell_1\ell_2)\ell\big)L;(\sigma_0(\sigma_1\sigma_2)s)\mathsf{S}\right\rangle$$

$$=\sum_\lambda\langle k\ell_0,n'_1\ell'_1\rangle\langle k'\ell'_0,n_1\ell_1;n'_2\ell'_2,n_2\ell_2\rangle\langle\sigma'_0(\sigma'_1\sigma'_2)|P_{01}|\sigma_0(\sigma_1\sigma_2)\rangle i^{\ell'_0+\ell'_1+\ell'_2-\ell_0-\ell_1-\ell_2}\times$$

$$\left\langle\left[Y^{[\ell'_0]}(\vec{r}_0)\left[Y^{[\ell'_1]}(\vec{r}_1)Y^{[\ell'_2]}(\vec{r}_2)\right]^{[\ell']}\right]^{[L]}\left\|\left[Y^{[\lambda]}(\vec{r}_0)Y^{[\lambda]}(\vec{r}_2)\right]^{[0]}\right\|\left[Y^{[\ell_0]}(\vec{r}_1)\left[Y^{[\ell_1]}(\vec{r}_0)Y^{[\ell_2]}(\vec{r}_2)\right]^{[\ell]}\right]^{[L]}\right\rangle$$

$$=\sum_{\lambda q}\langle k\ell_0,n'_1\ell'_1\rangle\langle k'\ell'_0,n_1\ell_1;n'_2\ell'_2,n_2\ell_2\rangle i^{\ell'_0+\ell'_1+\ell'_2-\ell_0-\ell_1-\ell_2}\times$$

$$\frac{\hat{\ell}_0}{\hat{L}}\delta_{\ell'_1\ell_0}[\lambda|\ell'_0|\ell_1][\lambda|\ell'_2|\ell_2]\begin{bmatrix}0 & \tfrac{1}{2} & \tfrac{1}{2} \\ \tfrac{1}{2} & \tfrac{1}{2} & s \\ \tfrac{1}{2} & s' & \mathsf{S}\end{bmatrix}\begin{bmatrix}0 & \ell_0 & \ell_0 \\ \ell_1 & \ell_2 & \ell \\ \ell_1 & q & L\end{bmatrix}\begin{bmatrix}\ell'_0 & \ell' & L \\ \ell_1 & q & L \\ \lambda & \lambda & 0\end{bmatrix}\begin{bmatrix}\ell'_1 & \ell'_2 & \ell' \\ \ell_0 & \ell_2 & q \\ 0 & \lambda & \lambda\end{bmatrix}\begin{bmatrix}\lambda & \lambda & 0 \\ \lambda & \lambda & 0 \\ 0 & 0 & 0\end{bmatrix},$$

(10)

$$\left\langle k'\bigl(\ell'_0(\ell'_1\ell'_2)\ell'\bigr)L;(\sigma'_0(\sigma'_1\sigma'_2)s')\mathsf{S}\left\|\frac{1}{r_{12}}P_{01}\right\|k\bigl(\ell_0(\ell_1\ell_2)\ell\bigr)L;(\sigma_0(\sigma_1\sigma_2)s)\mathsf{S}\right\rangle$$

$$=\sum_\lambda \langle k'\ell'_0,n_1\ell_1\rangle\langle n'\ell'_1,k\ell_0;n'_2\ell'_2,n_2\ell_2\rangle\langle \sigma'_0(\sigma'_1,\sigma'_2)|P_{01}|\sigma_0(\sigma_1\sigma_2)\rangle i^{\ell'_0+\ell'_1+\ell'_2-\ell_0-\ell_1-\ell_2}\times$$

$$\left\langle \left[Y^{[\ell'_0]}(\hat{r}_0)\bigl[Y^{[\ell'_1]}(\hat{r}_1)Y^{[\ell'_2]}(\hat{r}_2)\bigr]^{[\ell']}\right]^{[L]} \left\| \bigl[Y^{[\lambda]}(\hat{r}_1)Y^{[\lambda]}(\hat{r}_2)\bigr]^{[0]} \right\| \left[Y^{[\ell_0]}(\hat{r}_1)\bigl[Y^{[\ell_1]}(\hat{r}_0)Y^{[\ell_2]}(\hat{r}_2)\bigr]^{[\ell]}\right]^{[L]}\right\rangle$$

$$=\sum_\lambda \langle k'\ell'_0,n_1\ell_1\rangle\langle n'\ell'_1,k\ell_0;n'_2\ell'_2,n_2\ell_2\rangle i^{\ell'_0+\ell'_1+\ell'_2-\ell_0-\ell_1-\ell_2}\times$$

$$\frac{\hat{\ell}_1}{\hat{L}}\delta_{\ell_1\ell'_0}[\lambda|\ell'_1|\ell_0][\lambda|\ell'_2|\ell_2]\begin{bmatrix}0&\tfrac12&\tfrac12\\\tfrac12&\tfrac12&s\\\tfrac12&s'&\mathsf{S}\end{bmatrix}\begin{bmatrix}0&\ell_0&\ell_0\\\ell_1&\ell_2&\ell\\\ell_1&\ell'&L\end{bmatrix}\begin{bmatrix}\ell'_0&\ell'&L\\\ell_1&\ell'&L\\0&0&0\end{bmatrix}\begin{bmatrix}\ell'_1&\ell'_2&\ell'\\\ell_0&\ell_2&\ell'\\\lambda&\lambda&0\end{bmatrix}\begin{bmatrix}\lambda&\lambda&0\\\lambda&\lambda&0\\0&0&0\end{bmatrix},\quad(11)$$

$$\left\langle k'\bigl(\ell'_0(\ell'_1\ell'_2)\ell'\bigr)L;(\sigma'_0(\sigma'_1\sigma'_2)s')\mathsf{S}\|EP_{01}\|k\bigl(\ell_0(\ell_1\ell_2)\ell\bigr)L;(\sigma_0(\sigma_1\sigma_2)s)\mathsf{S}\right\rangle$$

$$=E\langle\varphi_{n'\ell'_1}|u_{\ell_0}(k)\rangle\langle u_{\ell'_0}(k')|\varphi_{n\ell_1}\rangle\langle\varphi_{n'\ell'_2}|\varphi_{n\ell_2}\rangle i^{\ell'_0+\ell'_1+\ell'_2-\ell_0-\ell_1-\ell_2}\langle\sigma'_0(\sigma'_1,\sigma'_2)|P_{01}|\sigma_0(\sigma_1\sigma_2)\rangle\times$$

$$\left\langle \left[Y^{[\ell'_0]}(\hat{r}_0)\bigl[Y^{[\ell'_1]}(\hat{r}_1)Y^{[\ell'_2]}(\hat{r}_2)\bigr]^{[\ell']}\right]^{[L]} \right\| \left[Y^{[\ell_0]}(\hat{r}_1)\bigl[Y^{[\ell_1]}(\hat{r}_0)Y^{[\ell_2]}(\hat{r}_2)\bigr]^{[\ell]}\right]^{[L]}\right\rangle$$

$$=E\langle\varphi_{n'\ell'_1}|u_{\ell_0}(k)\rangle\langle u_{\ell'_0}(k')|\varphi_{n\ell_1}\rangle\langle\varphi_{n'\ell'_2}|\varphi_{n\ell_2}\rangle(-1)^{\ell'_0+\ell'_1+\ell'_2-\ell_0-\ell_1-\ell_2}\begin{bmatrix}0&\tfrac12&\tfrac12\\\tfrac12&\tfrac12&s\\\tfrac12&s'&\mathsf{S}\end{bmatrix}\begin{bmatrix}0&\ell_0&\ell_0\\\ell_1&\ell_2&\ell\\\ell_1&\ell'&L\end{bmatrix}.\quad(12)$$

Minor simplification of Eqs. (8-12) gives rise to the same expressions presented in (Fursa and Bray, 1995), where the conventional algebraic approach was used. However, we emphasize that the invariant graphical method is so much simpler with the derivation completed by drawing the diagrams.

3.3 Electron-Li scattering (a four-electron system)

Derivations for these momentum-space potentials have been previously given for e-H (Bray and Stelbovics 1992), for e-Li (Fursa and Bray 1995), for e-Na (Bray, 1994), and for e-He (Bray, 1994), all using the conventional algebraic approach. The frozen core approximation was adopted for e-Li and e-Na in these studies, which could be the main reason for the small discrepancies between the theoretical calculations and experiments (Bray *et al.* 1999). With the enormous success of quantum scattering theories in describing scattering from one and two electron targets, one is naturally seeking to perform calculations with more complex systems and free of approximations. However, such an extension has been proven to be extremely tedious and in some cases even intractable. In the following, we derive a complete set of direct and exchange matrix elements for electron-Lithium scattering (previously not available) using the invariant graphical method. In this way, we are able to remove the frozen core approximation adopted by Bray *et al.* (1999, 1994) and Zhang et al. (1992). Again, the linear radial integral part of the direct and exchange elements is straightforward to work out and has the general form of

$$\langle n'_2 \ell'_2, n_2 \ell_2 \rangle \langle n'_3 \ell'_3, n_3 \ell_3 \rangle \langle k' \ell'_0, k\ell_0; n'_1 \ell'_1, n_1 \ell_1 \rangle$$

$$\equiv \langle \varphi_{n'_2 \ell'_2} | \varphi_{n_2 \ell_2} \rangle \langle \varphi_{n'_3 \ell'_3} | \varphi_{n_3 \ell_3} \rangle \int u_{\ell'_0}(k' r_0) u_{\ell_0}(k r_0) \varphi_{n' \ell'_1}(r_1) \varphi_{n \ell_1}(r_1) \frac{r_<^\lambda}{r_>^{\lambda+1}} dr_0 dr_1 .$$

The corresponding coupling and recoupling graphs are shown in Fig. 9-15 and the results are as the following,

$$\langle \vec{k}' \ell'_0 (\ell'_1 \ell'_2 \ell'_3) \ell' \| V_{01} \| \vec{k} \ell_0 (\ell_1 \ell_2 \ell_3) \ell \rangle$$

$$= \sum_\lambda \langle n'_2 \ell'_2, n_2 \ell_2 \rangle \langle n'_3 \ell'_3, n_3 \ell_3 \rangle \langle k' \ell'_0, k\ell_0; n'_1 \ell'_1, n_1 \ell_1 \rangle i^{\ell'_0 + \ell'_1 + \ell'_2 + \ell'_3 - \ell_0 - \ell_1 - \ell_2 - \ell_3} \times$$

$$\langle \sigma'_0 (\sigma'_1 (\sigma'_2 \sigma'_3)) | \sigma_0 (\sigma_1 (\sigma_2 \sigma_3)) \rangle \left\langle \left[ Y^{[\ell'_0]}(\vec{r}_0) \left[ Y^{[\ell'_1]}(\vec{r}_1) \left[ Y^{[\ell'_2]}(\vec{r}_2) Y^{[\ell'_3]}(\vec{r}_3) \right]^{[\ell'_{23}]} \right]^{[\ell']} \right]^{[L]} \right.$$

$$\left\| \left[ Y^{[\lambda]}(\vec{r}_0) Y^{[\lambda]}(\vec{r}_1) \right]^{[0]} \left\| \left[ Y^{[\ell_0]}(\vec{r}_0) \left[ Y^{[\ell_1]}(\vec{r}_1) \left[ Y^{[\ell_2]}(\vec{r}_2) Y^{[\ell_3]}(\vec{r}_3) \right]^{[\ell_{23}]} \right]^{[\ell]} \right]^{[L]} \right\rangle$$

$$= \sum_\lambda \langle n'_2 \ell'_2, n_2 \ell_2 \rangle \langle n'_3 \ell'_3, n_3 \ell_3 \rangle \langle k' \ell'_0, k\ell_0; n'_1 \ell'_1, n_1 \ell_1 \rangle i^{\ell'_0 + \ell'_1 + \ell'_2 + \ell'_3 - \ell_0 - \ell_1 - \ell_2 - \ell_3} \times$$

$$\frac{\hat{\ell}_2 \delta_{\ell_2 \ell'_2} \hat{\ell}_3 \delta_{\ell_3 \ell'_3}}{\hat{L}} [\lambda | \ell'_0 | \ell_0] [\lambda | \ell'_1 | \ell_1] \begin{bmatrix} \ell'_0 & \ell' & L \\ \ell_0 & \ell & L \\ \lambda & \lambda & 0 \end{bmatrix} \begin{bmatrix} \ell'_1 & \ell'_{23} & \ell' \\ \ell_1 & \ell_{23} & \ell \\ \lambda & 0 & \lambda \end{bmatrix} \begin{bmatrix} \lambda & \lambda & 0 \\ \lambda & 0 & \lambda \\ 0 & \lambda & \lambda \end{bmatrix} \begin{bmatrix} \ell'_2 & \ell'_3 & \ell'_{23} \\ \ell_2 & \ell_3 & \ell_{23} \\ 0 & 0 & 0 \end{bmatrix}$$

(13)

$$\langle \vec{k}' \ell'_0 (\ell'_1 \ell'_2 \ell'_3) \ell' \| V_{01} P_{01} \| \vec{k} \ell_0 (\ell_1 \ell_2 \ell_3) \ell \rangle$$

$$= \sum_\lambda \langle n'_2 \ell'_2, n_2 \ell_2 \rangle \langle n'_3 \ell'_3, n_3 \ell_3 \rangle \langle k' \ell'_0, n_1 \ell_1; k \ell_0, n'_1 \ell'_1 \rangle i^{\ell'_0 + \ell'_1 + \ell'_2 + \ell'_3 - \ell_0 - \ell_1 - \ell_2 - \ell_3}$$

$$\times \langle \sigma'_0 (\sigma'_1 (\sigma'_2 \sigma'_3)) | P_{01} | \sigma_0 (\sigma_1 (\sigma_2 \sigma_3)) \rangle \left\langle \left[ Y^{[\ell'_0]}(\vec{r}_0) \left[ Y^{[\ell'_1]}(\vec{r}_1) \left[ Y^{[\ell'_2]}(\vec{r}_2) Y^{[\ell'_3]}(\vec{r}_3) \right]^{[\ell'_{23}]} \right]^{[\ell']} \right]^{[L]} \right.$$

$$\left\| \left[ Y^{[\lambda]}(\vec{r}_0) Y^{[\lambda]}(\vec{r}_1) \right]^{[0]} \left\| \left[ Y^{[\ell_0]}(\vec{r}_1) \left[ Y^{[\ell_1]}(\vec{r}_0) \left[ Y^{[\ell_2]}(\vec{r}_2) Y^{[\ell_3]}(\vec{r}_3) \right]^{[\ell_{23}]} \right]^{[\ell]} \right]^{[L]} \right\rangle$$

$$= \sum_{\lambda q p} \langle n'_2 \ell'_2, n_2 \ell_2 \rangle \langle n'_3 \ell'_3, n_3 \ell_3 \rangle \langle k' \ell'_0, n_1 \ell_1; k \ell_0, n'_1 \ell'_1 \rangle i^{\ell'_0 + \ell'_1 + \ell'_2 + \ell'_3 - \ell_0 - \ell_1 - \ell_2 - \ell_3}$$

$$\times \frac{\hat{\ell}_2 \delta_{\ell_2 \ell'_2} \hat{\ell}_3 \delta_{\ell_3 \ell'_3}}{\hat{L}} [\lambda | \ell'_0 | \ell_1] [\lambda | \ell'_1 | \ell_0] \begin{bmatrix} 0 & \tfrac{1}{2} & \tfrac{1}{2} \\ \tfrac{1}{2} & p & s \\ \tfrac{1}{2} & s' & S \end{bmatrix} \begin{bmatrix} 0 & \ell_0 & \ell_0 \\ \ell_1 & \ell_{23} & \ell \\ \ell_1 & q & L \end{bmatrix} \begin{bmatrix} \ell'_0 & \ell' & L \\ \ell_1 & q & L \\ \lambda & \lambda & 0 \end{bmatrix}$$

$$\times \begin{bmatrix} \ell'_1 & \ell'_{23} & \ell' \\ \ell_0 & \ell_{23} & q \\ \lambda & 0 & \lambda \end{bmatrix} \begin{bmatrix} \lambda & \lambda & 0 \\ 0 & \lambda & \lambda \\ \lambda & \lambda & 0 \end{bmatrix} \begin{bmatrix} \ell'_2 & \ell'_3 & \ell'_{23} \\ \ell_2 & \ell_3 & \ell_{23} \\ 0 & 0 & 0 \end{bmatrix}$$

(14)

$$\left\langle \vec{k}'\ell'_0(\ell'_1\ell'_2\ell'_3)\ell' \right\| V_{02} P_{01} \left\| \vec{k}\ell_0(\ell_1\ell_2\ell_3)\ell \right\rangle$$

$$= \sum_\lambda \langle k\ell_0, n_1\ell'_1\rangle\langle n'_3\ell'_3, n_3\ell_3\rangle\langle k\ell'_0, n'_1\ell'_1; n'_2\ell'_2, n_2\ell_2\rangle i^{\ell'_0+\ell'_1+\ell'_2+\ell'_3-\ell_0-\ell_1-\ell_2-\ell_3}$$

$$\times \left\langle \sigma'_0(\sigma'_1(\sigma'_2\sigma'_3)) \middle| P_{01} \middle| \sigma_0(\sigma_1(\sigma_2\sigma_3)) \right\rangle \left\langle \left[ Y^{[\ell'_0]}(\vec{r}_0)\left[ Y^{[\ell'_1]}(\vec{r}_1)\left[ Y^{[\ell'_2]}(\vec{r}_2) Y^{[\ell'_3]}(\vec{r}_3)\right]^{[\ell'_{23}]}\right]^{[\ell']}\right]^{[L]}\right.$$

$$\left. \left\| \left[Y^{[\lambda]}(\vec{r}_0) Y^{[\lambda]}(\vec{r}_2)\right]^{[0]} \right\| \left[ Y^{[\ell_0]}(\vec{r}_1)\left[ Y^{[\ell_1]}(\vec{r}_0)\left[ Y^{[\ell_2]}(\vec{r}_2) Y^{[\ell_3]}(\vec{r}_3)\right]^{[\ell_{23}]}\right]^{[\ell]}\right]^{[L]} \right\rangle$$

$$= \sum_{\lambda qp} \langle k\ell_0, n_1\ell'_1\rangle\langle n'_3\ell'_3, n_3\ell_3\rangle\langle k\ell'_0, n'_1\ell'_1; n'_2\ell'_2, n_2\ell_2\rangle i^{\ell'_0+\ell'_1+\ell'_2+\ell'_3-\ell_0-\ell_1-\ell_2-\ell_3}$$

$$\times \frac{\hat{\ell}_0 \delta_{\ell_0\ell'_1} \hat{\ell}_3 \delta_{\ell_3\ell'_3}}{\hat{L}} [\lambda|\ell'_0|\ell_1][\lambda|\ell'_2|\ell_2] \begin{bmatrix} 0 & ½ & ½ \\ ½ & p & s \\ ½ & s' & S \end{bmatrix} \begin{bmatrix} 0 & \ell_0 & \ell_0 \\ \ell_1 & \ell_{23} & \ell \\ \ell_1 & q & L \end{bmatrix} \begin{bmatrix} \ell'_0 & \ell' & L \\ \ell_1 & q & L \\ \lambda & \lambda & 0 \end{bmatrix}$$

$$\times \begin{bmatrix} \ell'_1 & \ell'_{23} & \ell' \\ \ell_0 & \ell_{23} & q \\ 0 & \lambda & \lambda \end{bmatrix} \begin{bmatrix} \lambda & \lambda & 0 \\ \lambda & 0 & \lambda \\ 0 & \lambda & \lambda \end{bmatrix} \begin{bmatrix} \ell'_2 & \ell'_3 & \ell'_{23} \\ \ell_2 & \ell_3 & \ell_{23} \\ \lambda & 0 & \lambda \end{bmatrix}$$

(15)

$$\left\langle \vec{k}'\ell'_0(\ell'_1\ell'_2\ell'_3)\ell' \right\| V_{12} P_{01} \left\| \vec{k}\ell_0(\ell_1\ell_2\ell_3)\ell \right\rangle$$

$$= \sum_\lambda \langle k'\ell'_0, n_1\ell_1\rangle\langle n'_3\ell'_3, n_3\ell_3\rangle\langle n'_1\ell'_1, k\ell_0; n'_2\ell'_2, n_2\ell_2\rangle i^{\ell'_0+\ell'_1+\ell'_2+\ell'_3-\ell_0-\ell_1-\ell_2-\ell_3}$$

$$\times \left\langle \sigma'_0(\sigma'_1(\sigma'_2\sigma'_3)) \middle| P_{01} \middle| \sigma_0(\sigma_1(\sigma_2\sigma_3)) \right\rangle \left\langle \left[ Y^{[\ell'_0]}(\vec{r}_0)\left[ Y^{[\ell'_1]}(\vec{r}_1)\left[ Y^{[\ell'_2]}(\vec{r}_2) Y^{[\ell'_3]}(\vec{r}_3)\right]^{[\ell'_{23}]}\right]^{[\ell']}\right]^{[L]}\right.$$

$$\left. \left\| \left[Y^{[\lambda]}(\vec{r}_1) Y^{[\lambda]}(\vec{r}_2)\right]^{[0]} \right\| \left[ Y^{[\ell_0]}(\vec{r}_1)\left[ Y^{[\ell_1]}(\vec{r}_0)\left[ Y^{[\ell_2]}(\vec{r}_2) Y^{[\ell_3]}(\vec{r}_3)\right]^{[\ell_{23}]}\right]^{[\ell]}\right]^{[L]} \right\rangle$$

$$= \sum_{\lambda qp} \langle k'\ell'_0, n_1\ell_1\rangle\langle n'_3\ell'_3, n_3\ell_3\rangle\langle n'_1\ell'_1, k\ell_0; n'_2\ell'_2, n_2\ell_2\rangle i^{\ell'_0+\ell'_1+\ell'_2+\ell'_3-\ell_0-\ell_1-\ell_2-\ell_3}$$

$$\times \frac{\hat{\ell}_1 \delta_{\ell_1\ell'_0} \hat{\ell}_3 \delta_{\ell_3\ell'_3}}{\hat{L}} [\lambda|\ell'_1|\ell_0][\lambda|\ell'_2|\ell_2] \begin{bmatrix} 0 & ½ & ½ \\ ½ & p & s \\ ½ & s' & S \end{bmatrix} \begin{bmatrix} 0 & \ell_0 & \ell_0 \\ \ell_1 & \ell_{23} & \ell \\ \ell_1 & q & L \end{bmatrix} \begin{bmatrix} \ell'_0 & \ell' & L \\ \ell_1 & q & L \\ 0 & 0 & 0 \end{bmatrix}$$

$$\times \begin{bmatrix} \ell'_1 & \ell'_{23} & \ell' \\ \ell_0 & \ell_{23} & q \\ \lambda & \lambda & 0 \end{bmatrix} \begin{bmatrix} \lambda & \lambda & 0 \\ \lambda & \lambda & 0 \\ 0 & 0 & 0 \end{bmatrix} \begin{bmatrix} \ell'_2 & \ell'_3 & \ell'_{23} \\ \ell_2 & \ell_3 & \ell_{23} \\ \lambda & 0 & \lambda \end{bmatrix}$$

(16)

$$\langle \vec{k}'\ell'_0(\ell'_1\ell'_2\ell'_3)\ell'\|V_{23}P_{01}\|\vec{k}\ell_0(\ell_1\ell_2\ell_3)\ell\rangle$$

$$=\sum_{\lambda}\sum_{\lambda qp}\langle k'\ell'_0,n_1\ell_1\rangle\langle k\ell_0,n'_1\ell'_1\rangle\langle n'_2\ell'_2,n_2\ell_2;n'_3\ell'_3,n_3\ell_3\rangle i^{\ell'_0+\ell'_1+\ell'_2+\ell'_3-\ell_0-\ell_1-\ell_2-\ell_3}$$

$$\times\langle\sigma'_0(\sigma'_1(\sigma'_2\sigma'_3))|P_{01}|\sigma_0(\sigma_1(\sigma_2\sigma_3))\rangle\left\langle\left[Y^{[\ell'_0]}(\vec{r}_0)\left[Y^{[\ell'_1]}(\vec{r}_1)\left[Y^{[\ell'_2]}(\vec{r}_2)Y^{[\ell'_3]}(\vec{r}_3)\right]^{[\ell'_{23}]}\right]^{[\ell']}\right]^{[L]}\right.$$

$$\left.\left\|\left[Y^{[\lambda]}(\vec{r}_2)Y^{[\lambda]}(\vec{r}_3)\right]^{[0]}\right\|\left[Y^{[\ell_0]}(\vec{r}_1)\left[Y^{[\ell_a]}(\vec{r}_0)\left[Y^{[\ell_b]}(\vec{r}_2)Y^{[\ell_c]}(\vec{r}_3)\right]^{[\ell_{23}]}\right]^{[\ell]}\right]^{[L]}\right\rangle$$

$$=\sum_{\lambda qp}\langle k'\ell'_0,n_1\ell_1\rangle\langle k\ell_0,n'_1\ell'_1\rangle\langle n'_2\ell'_2,n_2\ell_2;n'_3\ell'_3,n_3\ell_3\rangle i^{\ell'_0+\ell'_1+\ell'_2+\ell'_3-\ell_0-\ell_1-\ell_2-\ell_3}$$

$$\times\frac{\hat{\ell}_1\delta_{\ell_1\ell'_0}\hat{\ell}_0\delta_{\ell_0\ell'_1}}{\hat{L}}[\lambda|\ell'_2|\ell_2][\lambda|\ell'_3|\ell_3]\begin{bmatrix}0 & \tfrac{1}{2} & \tfrac{1}{2}\\ \tfrac{1}{2} & p & s\\ \tfrac{1}{2} & s' & S\end{bmatrix}\begin{bmatrix}0 & \ell_0 & \ell_0\\ \ell_1 & \ell_{23} & \ell\\ \ell_1 & q & L\end{bmatrix}\begin{bmatrix}\ell'_0 & \ell' & L\\ \ell_1 & q & L\\ 0 & 0 & 0\end{bmatrix}$$

$$\times\begin{bmatrix}\ell'_1 & \ell'_{23} & \ell'\\ \ell_0 & \ell_{23} & q\\ 0 & 0 & 0\end{bmatrix}\begin{bmatrix}\ell'_2 & \ell'_3 & \ell'_{23}\\ \ell_2 & \ell_3 & \ell_{23}\\ \lambda & \lambda & 0\end{bmatrix}\begin{bmatrix}\lambda & \lambda & 0\\ \lambda & \lambda & 0\\ 0 & 0 & 0\end{bmatrix}$$

(17)

$$\langle \vec{k}'\ell'_0(\ell'_1\ell'_2\ell'_3)\ell'\|EP_{01}\|\vec{k}\ell_0(\ell_1\ell_2\ell_3)\ell\rangle$$

$$=E\langle\varphi_{n'_1\ell'_1}|u_{\ell_0}(k)\rangle\langle u_{\ell'_0}(k')|\varphi_{n_1\ell_1}\rangle\langle\varphi_{n'_2\ell'_2}|\varphi_{n_2\ell_2}\rangle\langle\varphi_{n'_3\ell'_3}|\varphi_{n_3\ell_3}\rangle i^{\ell'_0+\ell'_1+\ell'_2+\ell'_3-\ell_0-\ell_1-\ell_2-\ell_3}$$

$$\times\langle\sigma'_0(\sigma'_1(\sigma'_2\sigma'_3))|P_{01}|\sigma_0(\sigma_1(\sigma_2\sigma_3))\rangle\left\langle\left[Y^{[\ell'_0]}(\vec{r}_0)\left[Y^{[\ell'_1]}(\vec{r}_1)\left[Y^{[\ell'_2]}(\vec{r}_2)Y^{[\ell'_3]}(\vec{r}_3)\right]^{[\ell'_{23}]}\right]^{[\ell']}\right]^{[L]}\right\|$$

$$\left.\left[Y^{[\ell_0]}(\vec{r}_1)\left[Y^{[\ell_a]}(\vec{r}_0)\left[Y^{[\ell_b]}(\vec{r}_2)Y^{[\ell_c]}(\vec{r}_3)\right]^{[\ell_{23}]}\right]^{[\ell]}\right]^{[L]}\right\rangle$$

$$=(-1)^{\ell'_0+\ell'_1+\ell'_2+\ell'_3-\ell_0-\ell_1-\ell_2-\ell_3}E\langle\varphi_{n'_1\ell'_1}|u_{\ell_0}(k)\rangle\langle u_{\ell'_0}(k')|\varphi_{n_1\ell_1}\rangle\langle\varphi_{n'_2\ell'_2}|\varphi_{n_2\ell_2}\rangle\langle\varphi_{n'_3\ell'_3}|\varphi_{n_3\ell_3}\rangle$$

$$\times\sum_{pq}\begin{bmatrix}0 & \tfrac{1}{2} & \tfrac{1}{2}\\ \tfrac{1}{2} & p & s\\ \tfrac{1}{2} & s' & S\end{bmatrix}\begin{bmatrix}0 & \ell_0 & \ell_0\\ \ell_1 & \ell_{23} & \ell\\ \ell_1 & q & L\end{bmatrix}$$

(18)

4. Conclusions

The invariant graphical method developed by Danos and Fano (1998) is very powerful. The complex manipulation of angular momenta is reduced to the drawing of a compact graph, from which the final expressions can be readily read off. The procedure is transparent and simple to apply. It avoids writing out the intermediate expansions, yields immediately the selection rules for intermediate angular momenta, and helps finding the most direct and economical intermediate recoupling. It also requires a minimal use of rules in comparison with pages of formulas involved in other conventional methods.

**Acknowledgement**

JBW would like to thank H.Y. Wu for cross checking the phases in Eq. 8 using the conventional algebraic method. The authors also acknowledge support from The University of Western Australia, Murdoch University, and Curtin University of Technology.

**Figure captions**

Figure 1. (a) Basic recoupling transformation graph; (b) Recoupling of three tensors.

Figure 2. Recoupling graph for Eq. 5.

Figure 3. Recoupling graph of the spin tensors for Eqs. 8-12.

Figure 4. Recoupling graph of the orbit tensors for Eq. 8

Figure 5. Recoupling graph of the orbit tensors for Eq. 9

Figure 6. Recoupling graph of the orbit tensors for Eq. 10

Figure 7. Recoupling graph of the orbit tensors for Eq. 11

Figure 8. Recoupling graph of the orbit tensors for Eq. 12

Figure 9. Recoupling graph of the spin tensors for Eqs. 13-18.

Figure 10. Recoupling graph of the orbit tensors for Eq. 13

Figure 11. Recoupling graph of the orbit tensors for Eq. 14

Figure 12. Recoupling graph of the orbit tensors for Eq. 15

Figure 13. Recoupling graph of the orbit tensors for Eq. 16

Figure 14. Recoupling graph of the orbit tensors for Eq. 17

Figure 15. Recoupling graph of the orbit tensors for Eq. 18

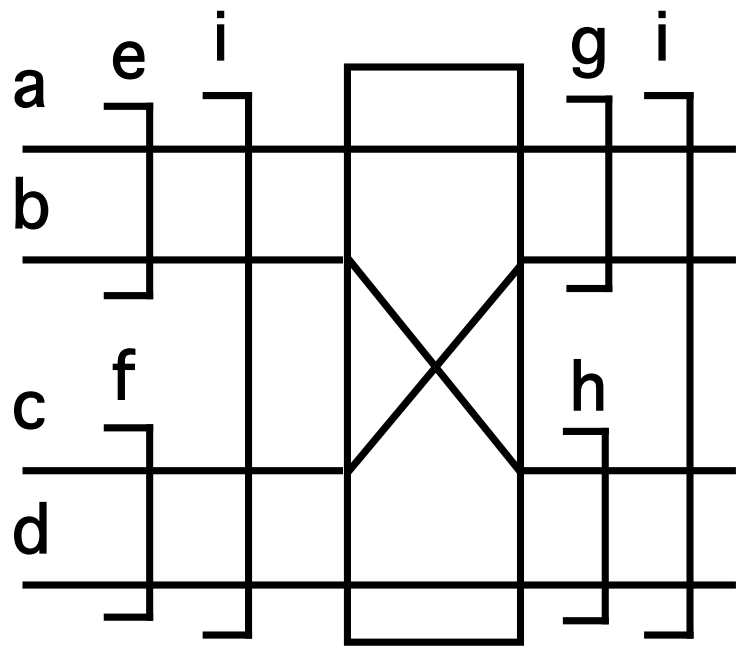 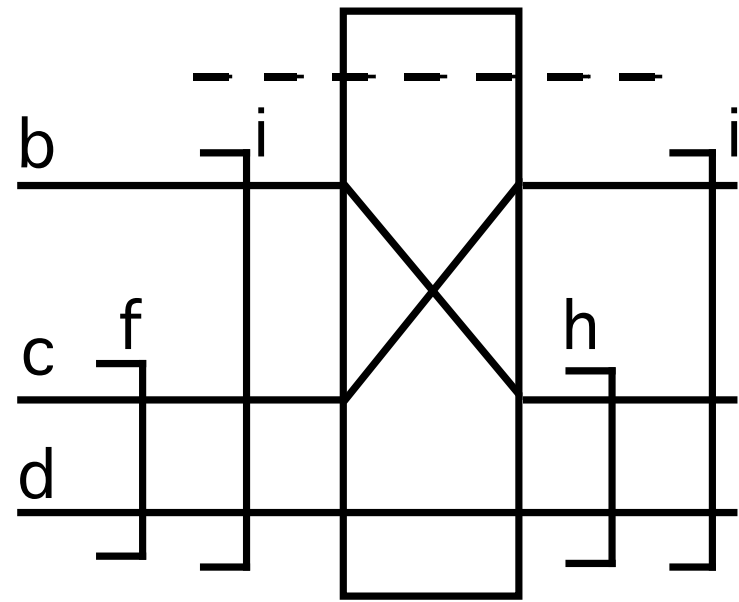

Figure 1

$$\left\langle \left[ Y^{[\ell'_a]} Y^{[\ell'_b]} \right]^{[\ell]} \middle\| \left[ Y^{[\lambda]} Y^{[\lambda]} \right]^{[0]} \middle\| \left[ Y^{[\ell_a]} Y^{[\ell_b]} \right]^{[\ell]} \right\rangle$$

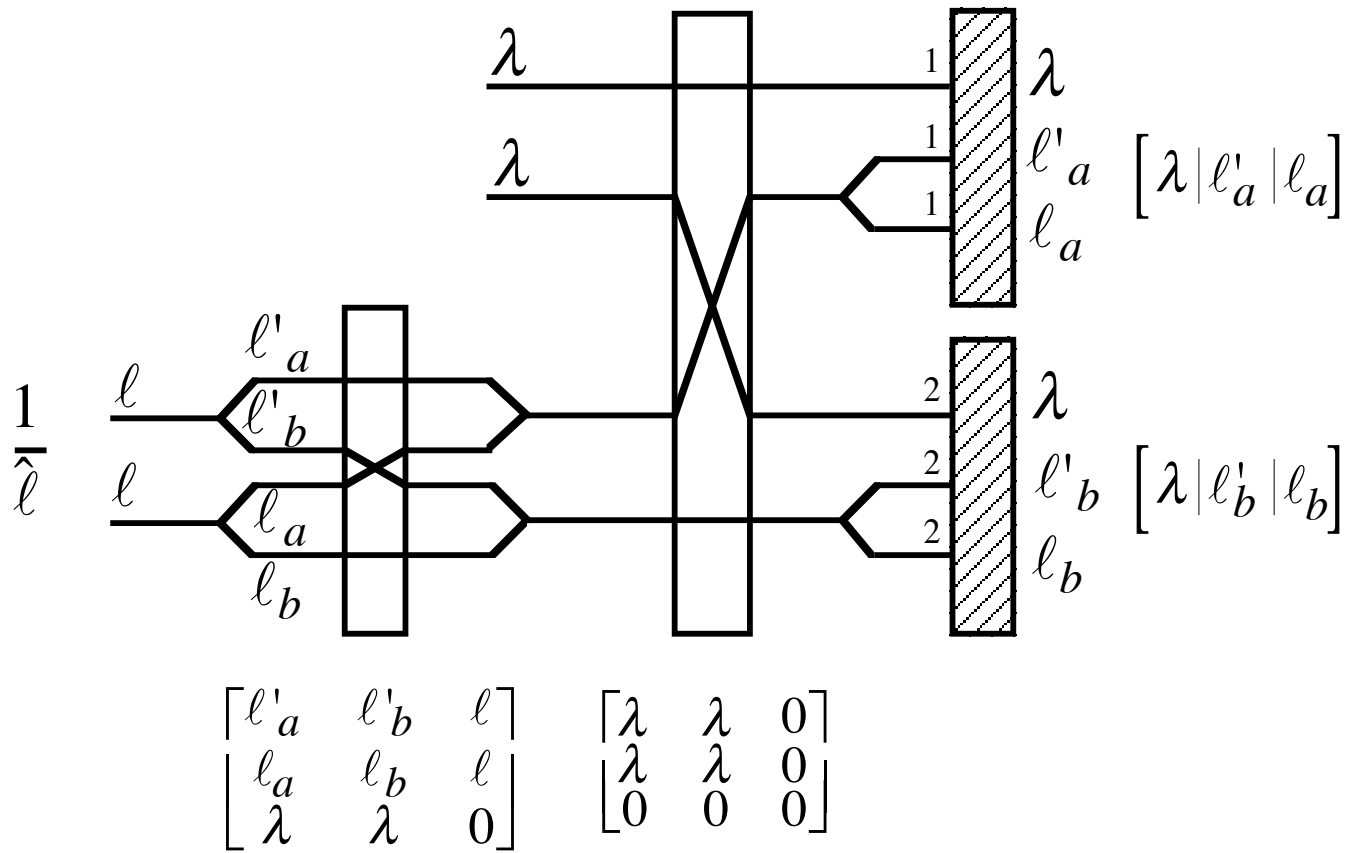

Figure 2

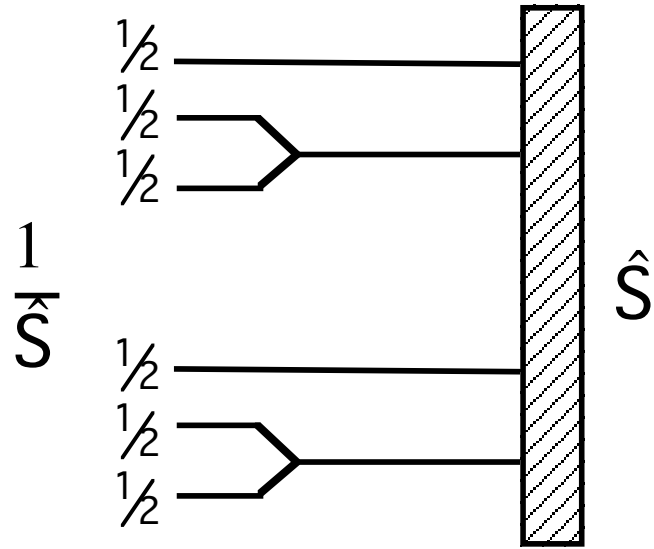 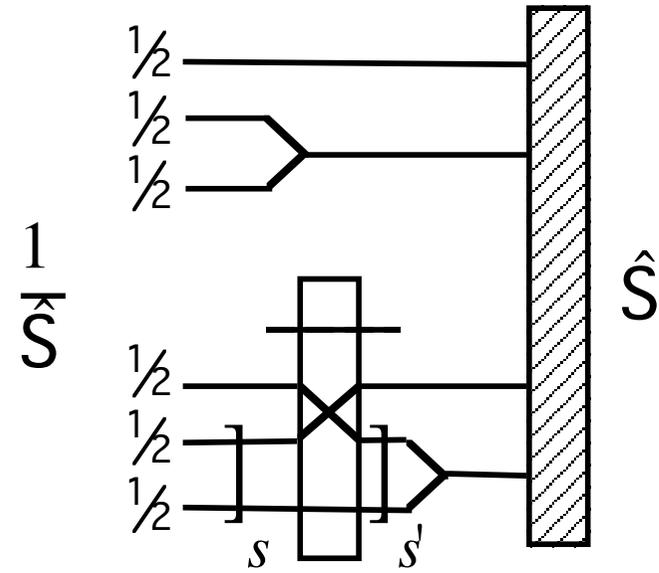

Figure 3

$$\left\langle \left[ Y^{[\ell'_0]}(\vec{r}_0) \left[ Y^{[\ell'_1]}(\vec{r}_1) Y^{[\ell'_2]}(\vec{r}_2) \right]^{[\ell']} \right]^{[L]} \middle\| \left[ Y^{[\lambda]}(\vec{r}_0) Y^{[\lambda]}(\vec{r}_1) \right]^{[0]} \middle\| \left[ Y^{[\ell_0]}(\vec{r}_0) \left[ Y^{[\ell_1]}(\vec{r}_1) Y^{[\ell_2]}(\vec{r}_2) \right]^{[\ell]} \right]^{[L]} \right\rangle$$

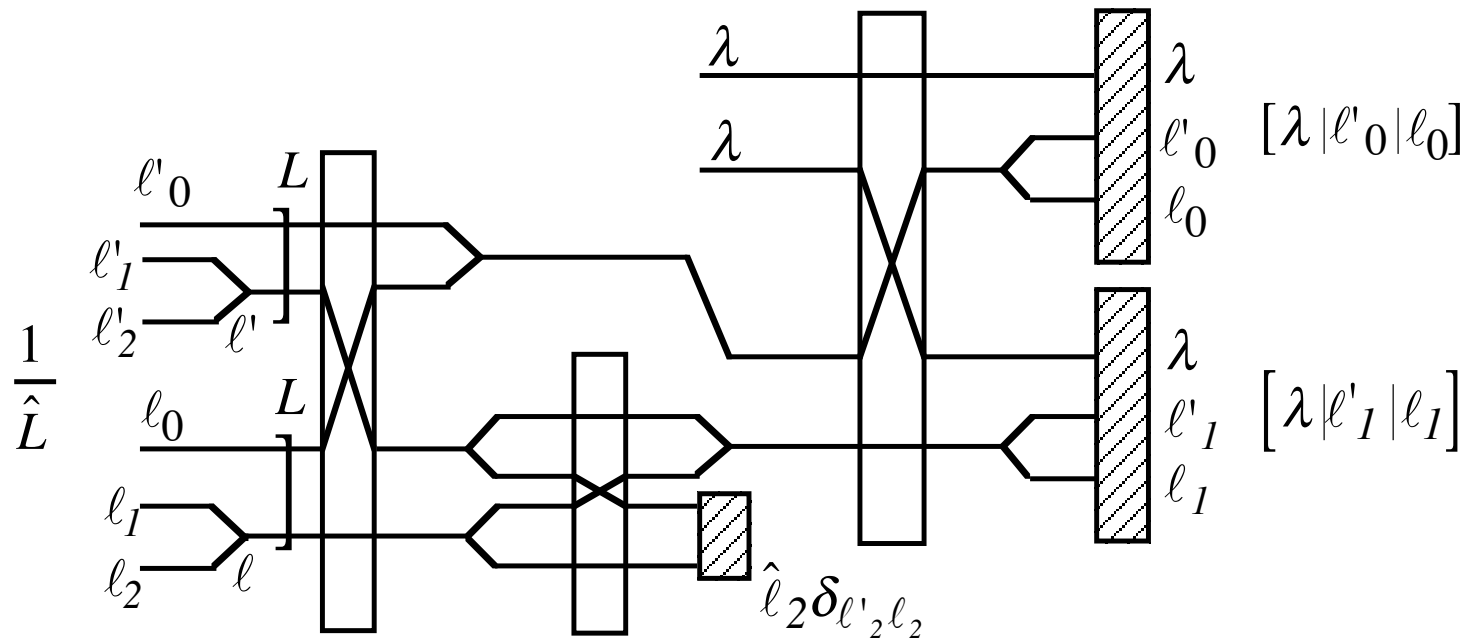

$$\frac{1}{\hat{L}} \begin{bmatrix} \ell'_0 & \ell' & L \\ \ell_0 & \ell & L \\ \lambda & \lambda & 0 \end{bmatrix} \begin{bmatrix} \ell'_1 & \ell'_2 & \ell' \\ \ell_1 & \ell_2 & \ell \\ \lambda & 0 & \lambda \end{bmatrix} \begin{bmatrix} \lambda & \lambda & 0 \\ \lambda & \lambda & 0 \\ 0 & 0 & 0 \end{bmatrix}$$

Figure 4

$$\left\langle \left[ Y^{[\ell'_0]}(\vec{r}_0) \left[ Y^{[\ell'_1]}(\vec{r}_1) Y^{[\ell'_2]}(\vec{r}_2) \right]^{[\ell']} \right]^{[L]} \middle\| \left[ Y^{[\lambda]}(\vec{r}_0) Y^{[\lambda]}(\vec{r}_1) \right]^{[0]} \middle\| \left[ Y^{[\ell_0]}(\vec{r}_1) \left[ Y^{[\ell_1]}(\vec{r}_0) Y^{[\ell_2]}(\vec{r}_2) \right]^{[\ell]} \right]^{[L]} \right\rangle$$

$$\frac{1}{\hat{L}} \begin{bmatrix} 0 & \ell_0 & \ell_0 \\ \ell_1 & \ell_2 & \ell \\ \ell_1 & q & L \end{bmatrix} \begin{bmatrix} \ell'_0 & \ell' & L \\ \ell_1 & q & L \\ \lambda & \lambda & 0 \end{bmatrix} \begin{bmatrix} \ell'_1 & \ell'_2 & \ell' \\ \ell_0 & \ell_2 & q \\ \lambda & 0 & \lambda \end{bmatrix} \begin{bmatrix} \lambda & \lambda & 0 \\ \lambda & \lambda & 0 \\ 0 & 0 & 0 \end{bmatrix}$$

Figure 5

$$\left\langle \left[Y^{[\ell'_0]}(\vec{r}_0)\left[Y^{[\ell'_1]}(\vec{r}_1)Y^{[\ell'_2]}(\vec{r}_2)\right]^{[\ell']}\right]^{[L]} \middle\| \left[Y^{[\lambda]}(\vec{r}_0)Y^{[\lambda]}(\vec{r}_2)\right]^{[0]} \middle\| \left[Y^{[\ell_0]}(\vec{r}_1)\left[Y^{[\ell_1]}(\vec{r}_0)Y^{[\ell_2]}(\vec{r}_2)\right]^{[\ell]}\right]^{[L]} \right\rangle$$

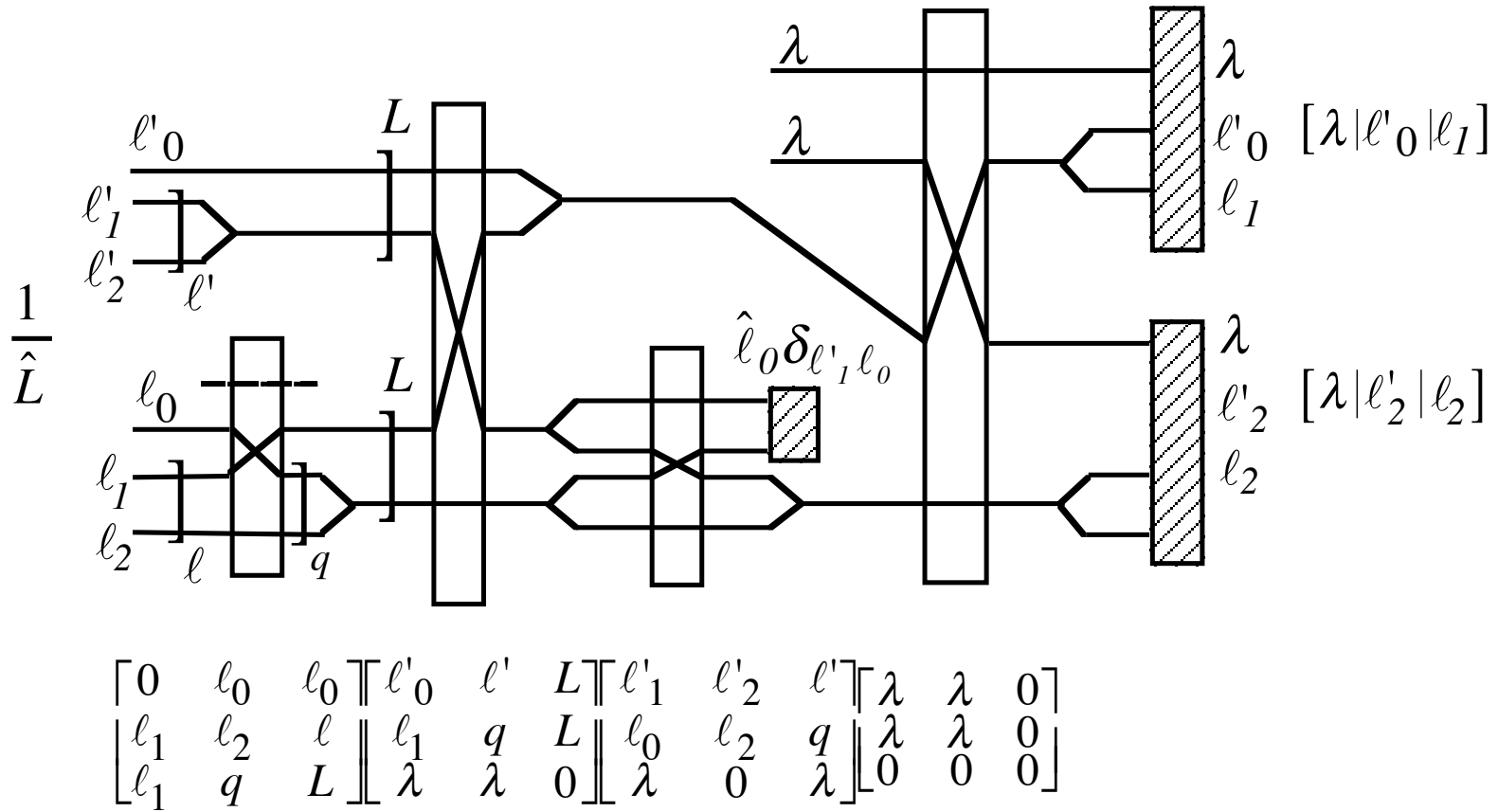

$$\begin{bmatrix} 0 & \ell_0 & \ell_0 \\ \ell_1 & \ell_2 & \ell \\ \ell_1 & q & L \end{bmatrix} \begin{bmatrix} \ell'_0 & \ell' & L \\ \ell_1 & q & L \\ \lambda & \lambda & 0 \end{bmatrix} \begin{bmatrix} \ell'_1 & \ell'_2 & \ell' \\ \ell_0 & \ell_2 & q \\ \lambda & 0 & \lambda \end{bmatrix} \begin{bmatrix} \lambda & \lambda & 0 \\ \lambda & \lambda & 0 \\ 0 & 0 & 0 \end{bmatrix}$$

Figure 6

$$\left\langle \left[ Y^{[\ell'_0]}(\vec{r}_0) \left[ Y^{[\ell'_1]}(\vec{r}_1) Y^{[\ell'_2]}(\vec{r}_2) \right]^{[\ell']} \right]^{[L]} \middle\| \left[ Y^{[\lambda]}(\vec{r}_1) Y^{[\lambda]}(\vec{r}_2) \right]^{[0]} \middle\| \left[ Y^{[\ell_0]}(\vec{r}_1) \left[ Y^{[\ell_1]}(\vec{r}_0) Y^{[\ell_2]}(\vec{r}_2) \right]^{[\ell]} \right]^{[L]} \right\rangle$$

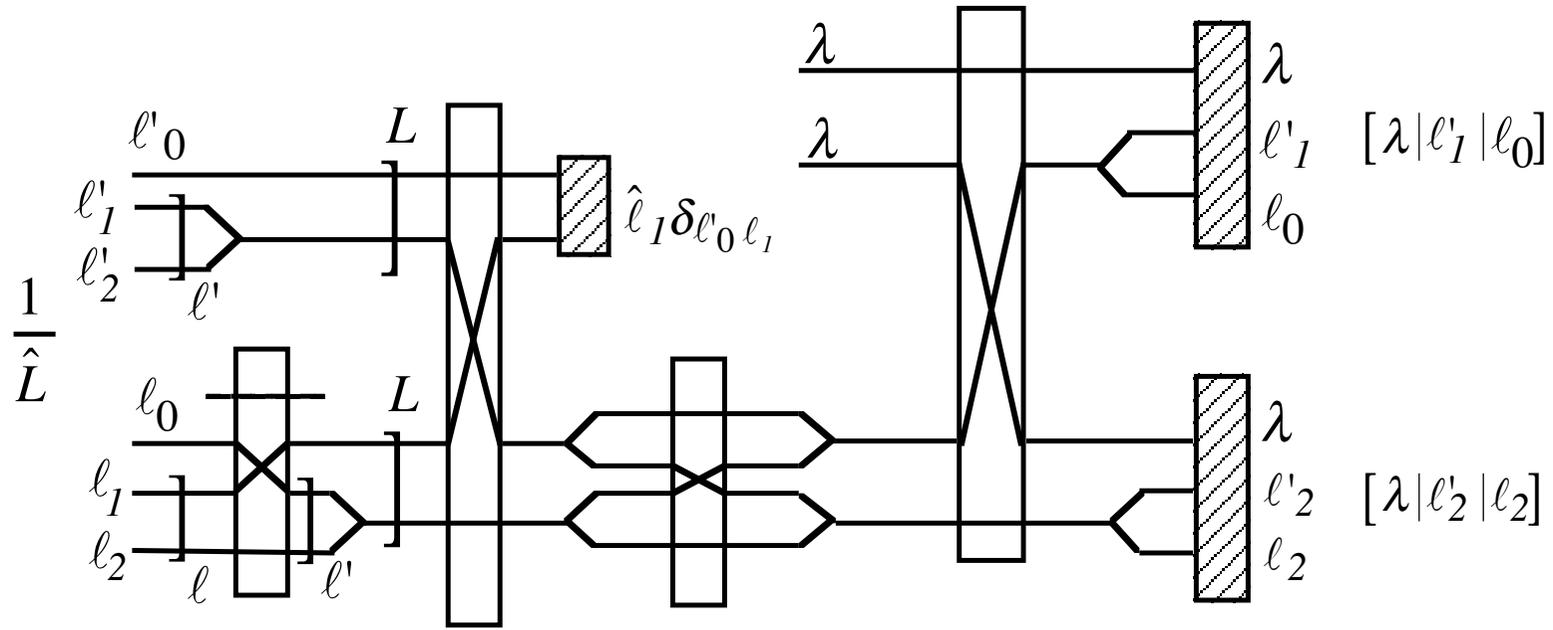

$$\frac{1}{\hat{L}} \begin{bmatrix} 0 & \ell_0 & \ell_0 \\ \ell_1 & \ell_2 & \ell \\ \ell_1 & \ell & L \end{bmatrix} \begin{bmatrix} \ell'_0 & \ell' & L \\ \ell_1 & \ell' & L \\ 0 & 0 & 0 \end{bmatrix} \begin{bmatrix} \ell'_1 & \ell'_2 & \ell' \\ \ell_0 & \ell_2 & \ell' \\ \lambda & \lambda & 0 \end{bmatrix} \begin{bmatrix} \lambda & \lambda & 0 \\ \lambda & \lambda & 0 \\ 0 & 0 & 0 \end{bmatrix}$$

Figure 7

$$\left\langle \left[ Y^{[\ell'_0]}(\vec{r_0}) \left[ Y^{[\ell'_1]}(\vec{r_1}) Y^{[\ell'_2]}(\vec{r_2}) \right]^{[\ell']} \right]^{[L]} \middle\| \left[ Y^{[\ell_0]}(\vec{r_1}) \left[ Y^{[\ell_1]}(\vec{r_0}) Y^{[\ell_2]}(\vec{r_2}) \right]^{[\ell]} \right]^{[L]} \right\rangle$$

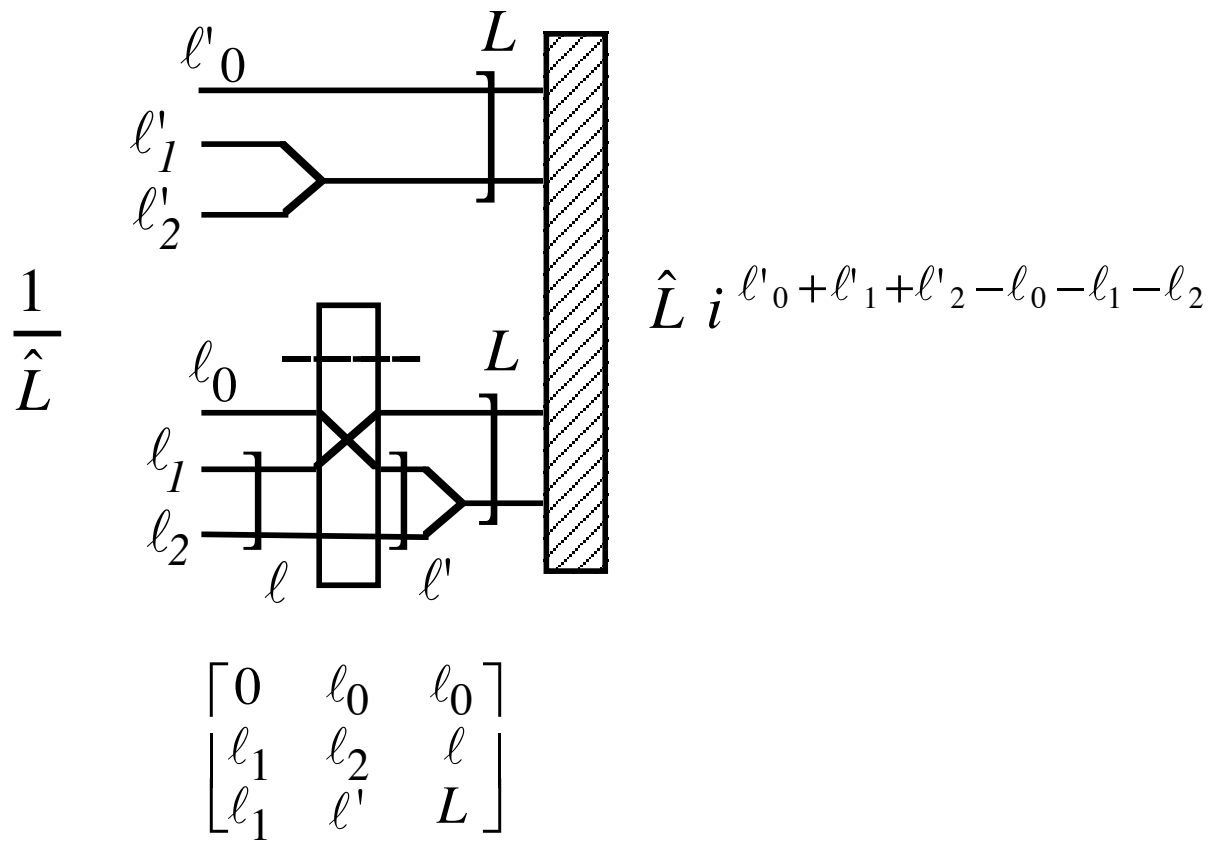

$$\frac{1}{\hat{L}} \qquad \hat{L}\, i^{\ell'_0 + \ell'_1 + \ell'_2 - \ell_0 - \ell_1 - \ell_2}$$

$$\begin{bmatrix} 0 & \ell_0 & \ell_0 \\ \ell_1 & \ell_2 & \ell \\ \ell_1 & \ell' & L \end{bmatrix}$$

Figure 8

$$\langle \sigma'_0 (\sigma'_1 \sigma'_2 \sigma'_3) | \sigma_0(\sigma_1\sigma_2\sigma_3) \rangle \qquad \langle \sigma'_0 (\sigma'_1 \sigma'_2 \sigma'_3) | \sigma_1(\sigma_0\sigma_2\sigma_3) \rangle$$

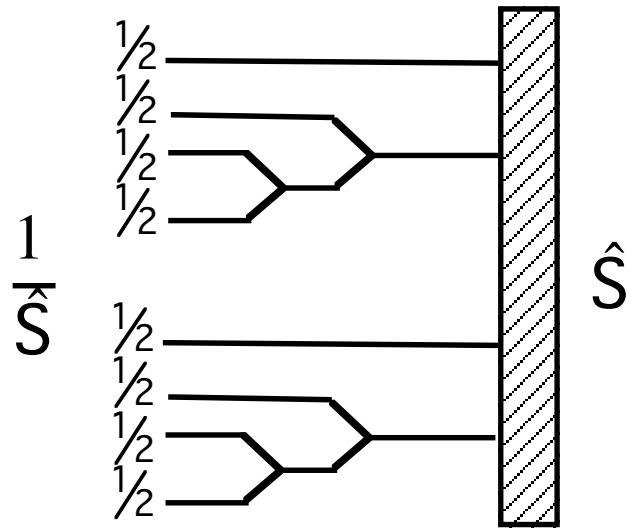
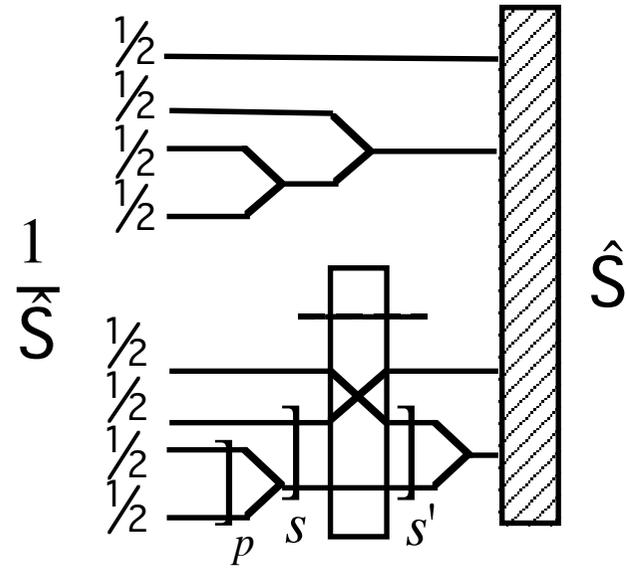

$$\begin{bmatrix} 0 & \tfrac{1}{2} & \tfrac{1}{2} \\ \tfrac{1}{2} & p & s \\ \tfrac{1}{2} & s' & S \end{bmatrix}$$

Figure 9

$$\left\langle \left[Y^{[\ell'_0]}(\vec{r}_0)\left[Y^{[\ell'_1]}(\vec{r}_1)\left[Y^{[\ell'_2]}(\vec{r}_2)Y^{[\ell'_3]}(\vec{r}_3)\right]^{[\ell'_{23}]}\right]^{[\ell']}\right]^{[L]} \middle\| \left[Y^{[\lambda]}(\vec{r}_0)Y^{[\lambda]}(\vec{r}_1)\right]^{[0]} \middle\|\right.$$

$$\left. \left[Y^{[\ell_0]}(\vec{r}_0)\left[Y^{[\ell_1]}(\vec{r}_1)\left[Y^{[\ell_2]}(\vec{r}_2)Y^{[\ell_3]}(\vec{r}_3)\right]^{[\ell_{23}]}\right]^{[\ell]}\right]^{[L]} \right\rangle$$

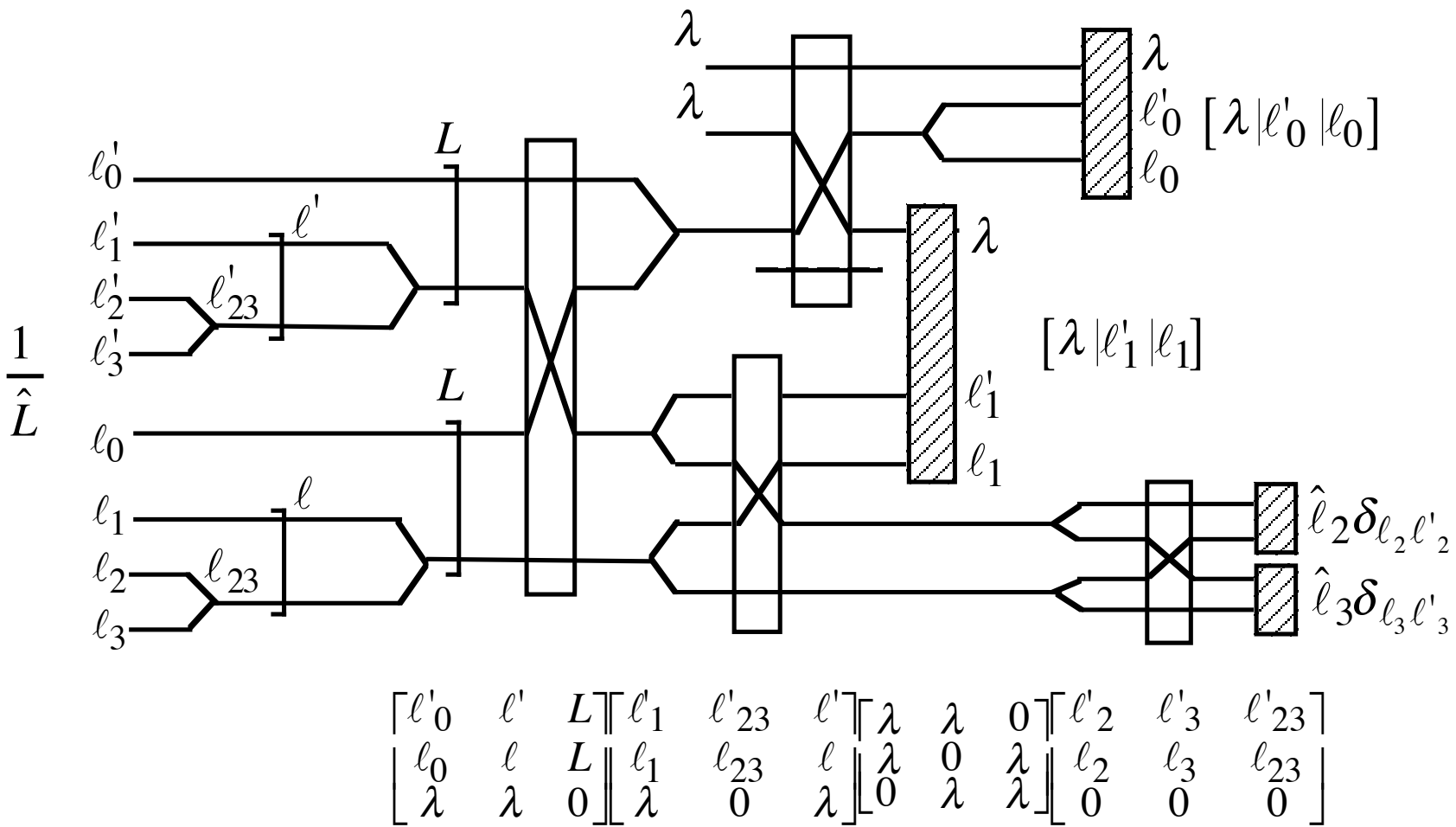

Figure 10

$$\left\langle \left[ Y^{[\ell_0]}(\vec{r}_0) \left[ Y^{[\ell_1]}(\vec{r}_1) \left[ Y^{[\ell_2]}(\vec{r}_2) Y^{[\ell_3]}(\vec{r}_3) \right]^{[\ell'_{23}]} \right]^{[\ell']} \right]^{[L]} \middle\| \left[ Y^{[\lambda]}(\vec{r}_0) Y^{[\lambda]}(\vec{r}_1) \right]^{[0]} \middle\| \right.$$

$$\left. \left[ Y^{[\ell_0]}(\vec{r}_1) \left[ Y^{[\ell_1]}(\vec{r}_0) \left[ Y^{[\ell_2]}(\vec{r}_2) Y^{[\ell_3]}(\vec{r}_3) \right]^{[\ell_{23}]} \right]^{[\ell]} \right]^{[L]} \right\rangle$$

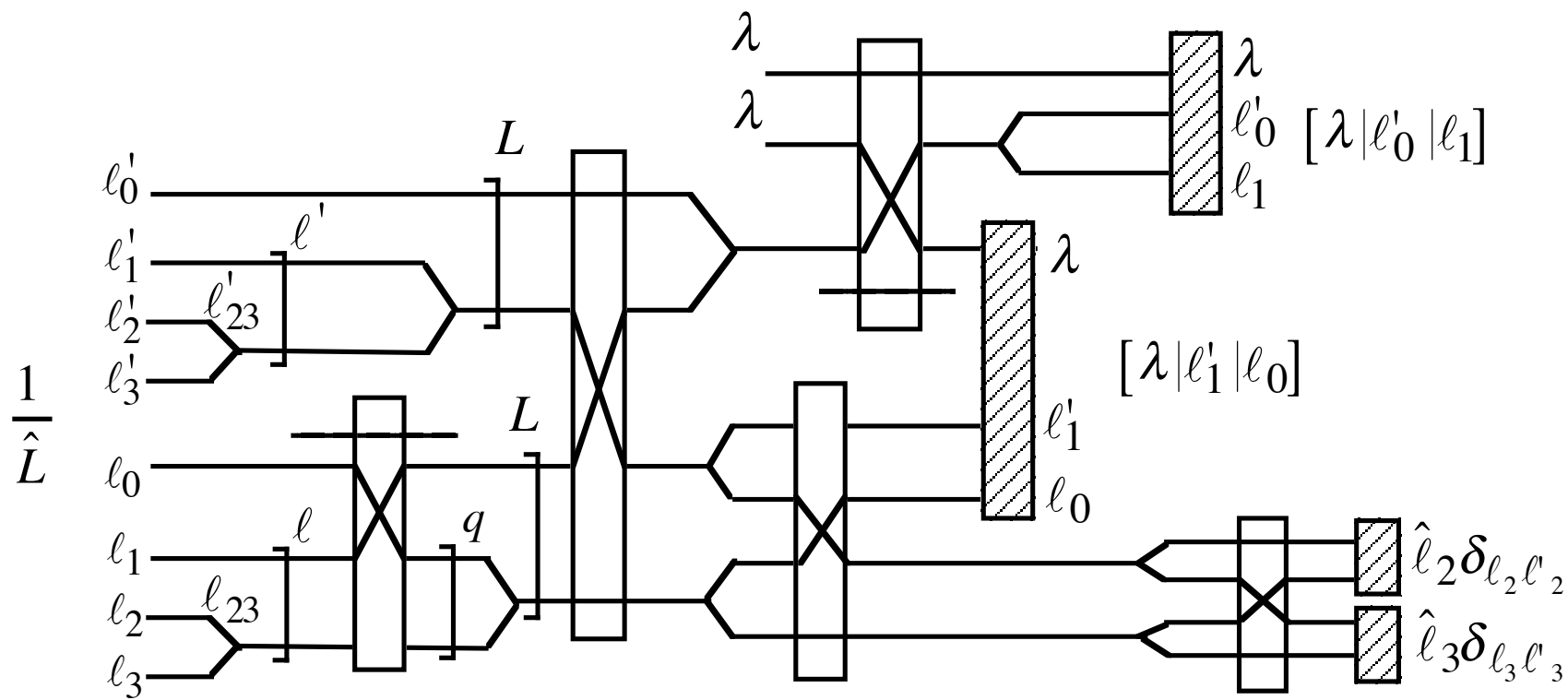

Figure 11

$$\left\langle \left[ Y^{[\ell_0]}(\vec{r}_0) \left[ Y^{[\ell_1]}(\vec{r}_1) \left[ Y^{[\ell_2]}(\vec{r}_2) Y^{[\ell'_3]}(\vec{r}_3) \right]^{[\ell'_{23}]} \right]^{[\ell']} \right]^{[L]} \middle\| \left[ Y^{[\lambda]}(\vec{r}_0) Y^{[\lambda]}(\vec{r}_2) \right]^{[0]} \right\| $$

$$\left. \left[ Y^{[\ell_0]}(\vec{r}_1) \left[ Y^{[\ell_1]}(\vec{r}_0) \left[ Y^{[\ell_2]}(\vec{r}_2) Y^{[\ell_3]}(\vec{r}_3) \right]^{[\ell_{23}]} \right]^{[\ell]} \right]^{[L]} \right\rangle$$

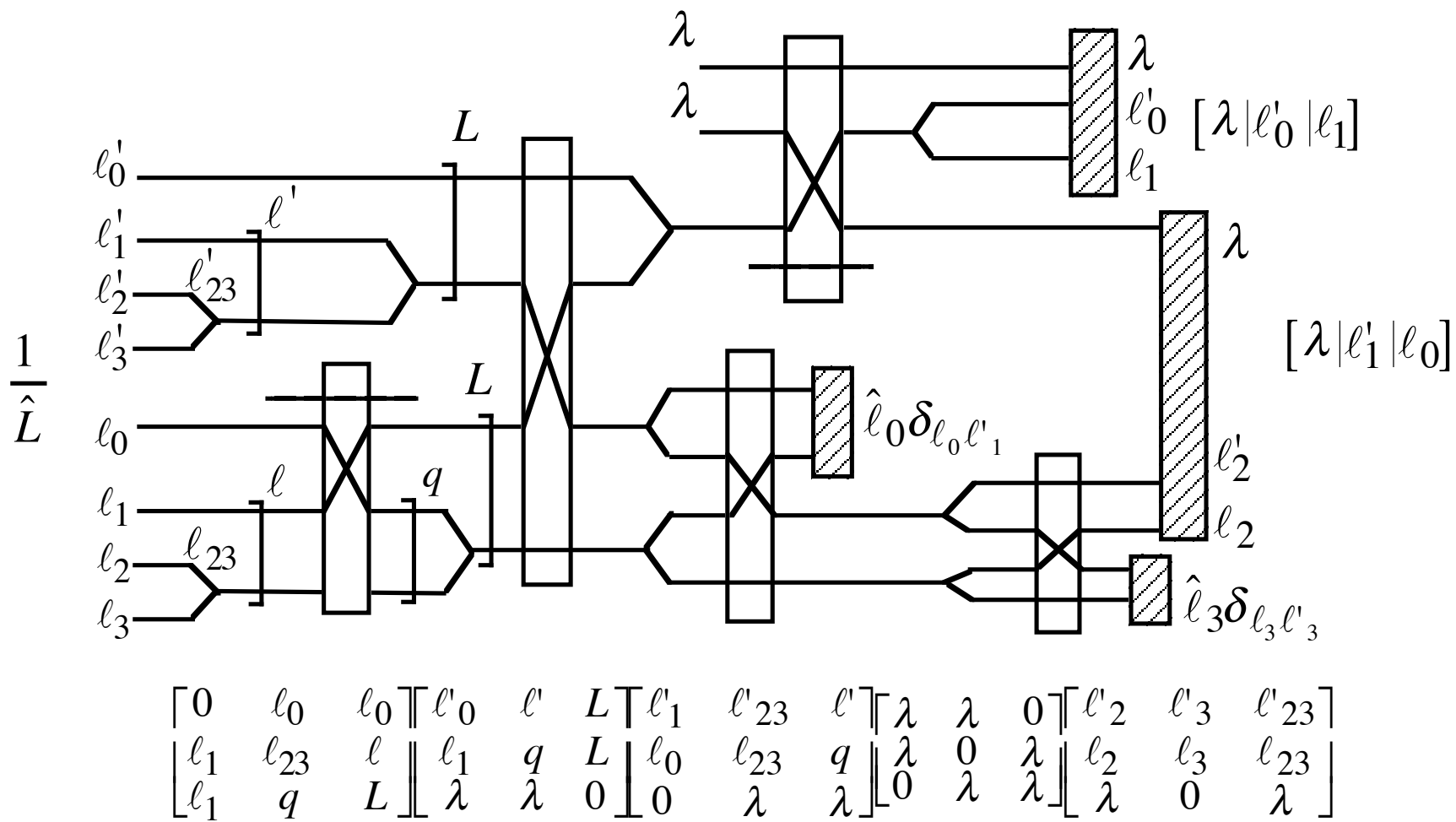

$$\frac{1}{\hat{L}} \begin{bmatrix} 0 & \ell_0 & \ell_0 \\ \ell_1 & \ell_{23} & \ell \\ \ell_1 & q & L \end{bmatrix} \begin{bmatrix} \ell'_0 & \ell' & L \\ \ell_1 & q & L \\ \lambda & \lambda & 0 \end{bmatrix} \begin{bmatrix} \ell'_1 & \ell'_{23} & \ell' \\ \ell_0 & \ell_{23} & q \\ 0 & \lambda & \lambda \end{bmatrix} \begin{bmatrix} \lambda & \lambda & 0 \\ \lambda & 0 & \lambda \\ 0 & \lambda & \lambda \end{bmatrix} \begin{bmatrix} \ell'_2 & \ell'_3 & \ell'_{23} \\ \ell_2 & \ell_3 & \ell_{23} \\ \lambda & 0 & \lambda \end{bmatrix}$$

Figure 12

$$\left\langle \left[ Y^{[\ell'_0]}(\vec{r}_0) \left[ Y^{[\ell'_1]}(\vec{r}_1) \left[ Y^{[\ell'_2]}(\vec{r}_2) Y^{[\ell'_3]}(\vec{r}_3) \right]^{[\ell'_{23}]} \right]^{[\ell']} \right]^{[L]} \middle\| \left[ Y^{[\lambda]}(\vec{r}_1) Y^{[\lambda]}(\vec{r}_2) \right]^{[0]} \middle\| \right.$$

$$\left. \left[ Y^{[\ell_0]}(\vec{r}_1) \left[ Y^{[\ell_1]}(\vec{r}_0) \left[ Y^{[\ell_2]}(\vec{r}_2) Y^{[\ell_3]}(\vec{r}_3) \right]^{[\ell_{23}]} \right]^{[\ell]} \right]^{[L]} \right\rangle$$

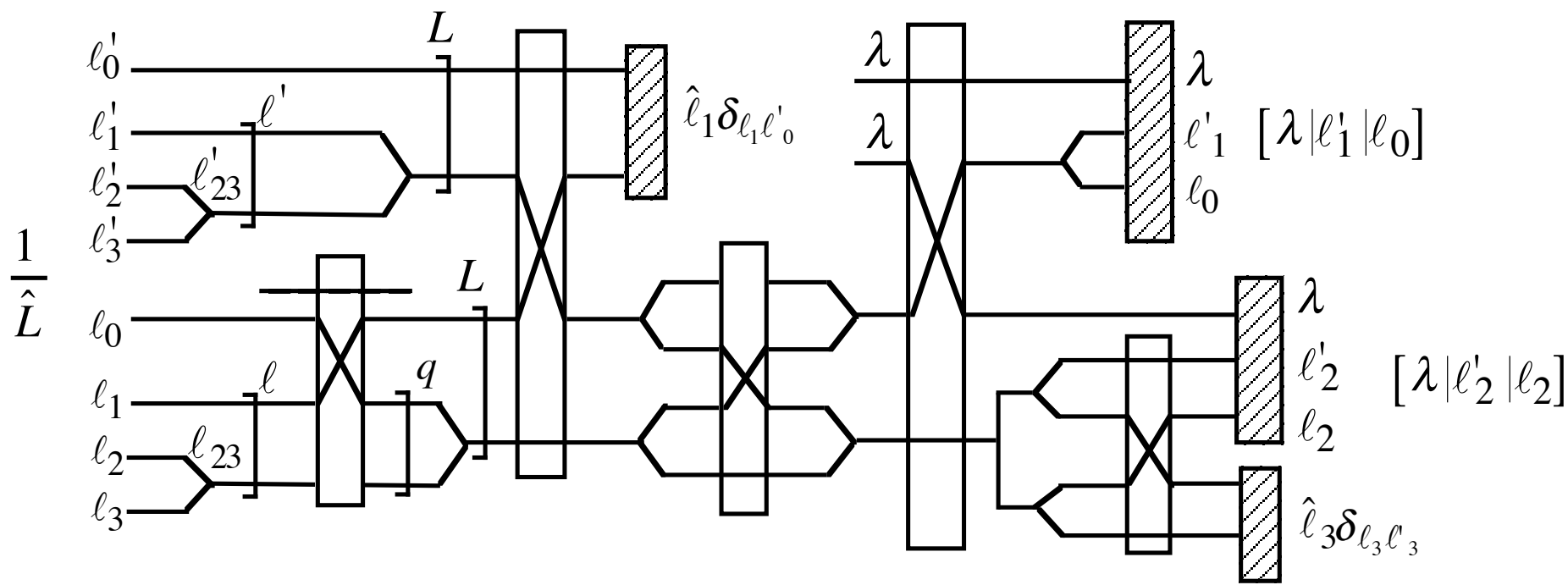

Figure 13

$$\left\langle \left[ Y^{[\ell_0]}(\vec{r}_0) \left[ Y^{[\ell'_1]}(\vec{r}_1) \left[ Y^{[\ell'_2]}(\vec{r}_2) Y^{[\ell'_3]}(\vec{r}_3) \right]^{[\ell'_{23}]} \right]^{[\ell']} \right]^{[L]} \middle\| \left[ Y^{[\lambda]}(\vec{r}_2) Y^{[\lambda]}(\vec{r}_3) \right]^{[0]} \middle\| \right.$$

$$\left. \left[ Y^{[\ell_0]}(\vec{r}_1) \left[ Y^{[\ell_a]}(\vec{r}_0) \left[ Y^{[\ell_b]}(\vec{r}_2) Y^{[\ell_c]}(\vec{r}_3) \right]^{[\ell_{23}]} \right]^{[\ell]} \right]^{[L]} \right\rangle$$

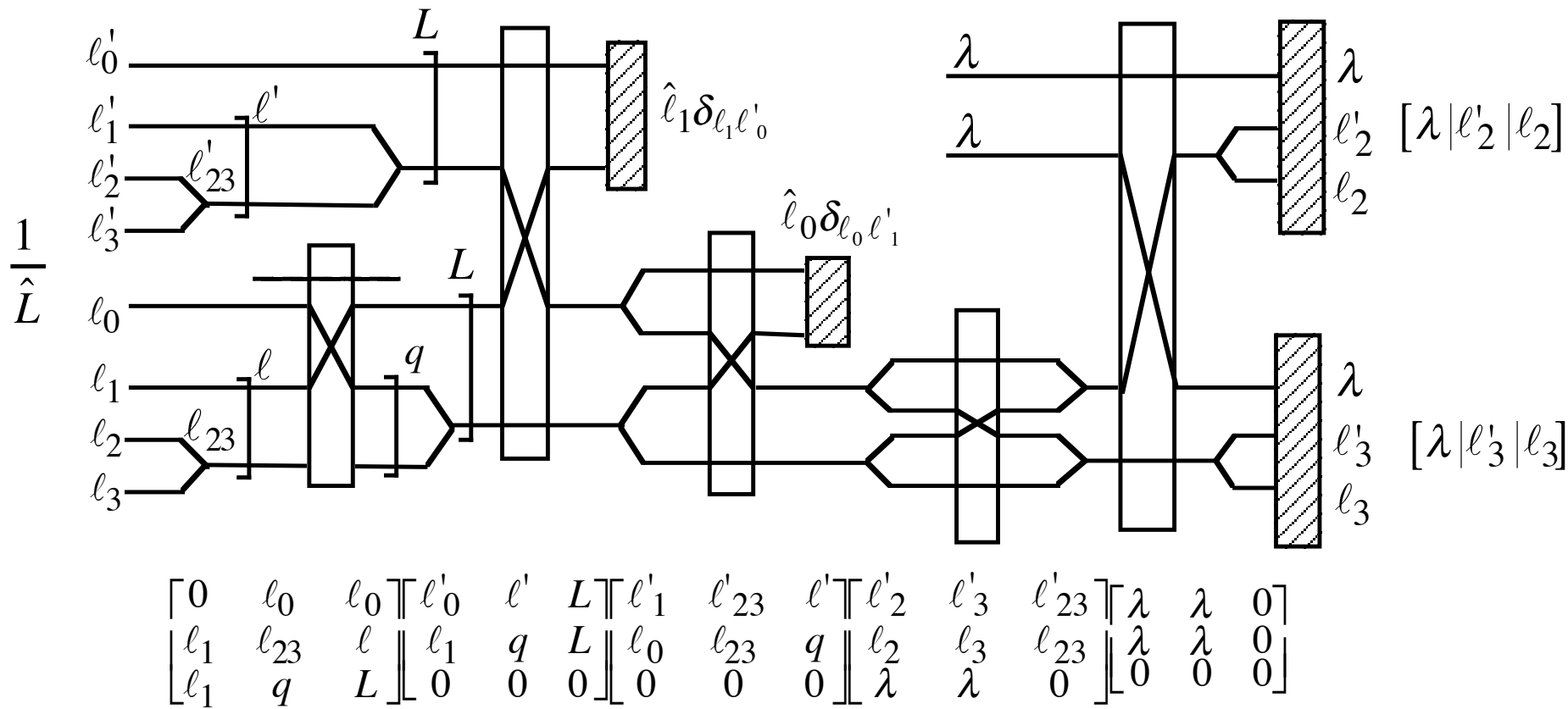

Figure 14

$$\left\langle \left[ Y^{[\ell'_0]}(\vec{r}_0) \left[ Y^{[\ell'_1]}(\vec{r}_1) \left[ Y^{[\ell'_2]}(\vec{r}_2) Y^{[\ell'_3]}(\vec{r}_3) \right]^{[\ell'_{23}]} \right]^{[\ell']} \right]^{[L]} \middle\| \left[ Y^{[\ell_0]}(\vec{r}_1) \left[ Y^{[\ell_a]}(\vec{r}_0) \left[ Y^{[\ell_b]}(\vec{r}_2) Y^{[\ell_c]}(\vec{r}_3) \right]^{[\ell_{23}]} \right]^{[\ell]} \right]^{[L]} \right\rangle$$

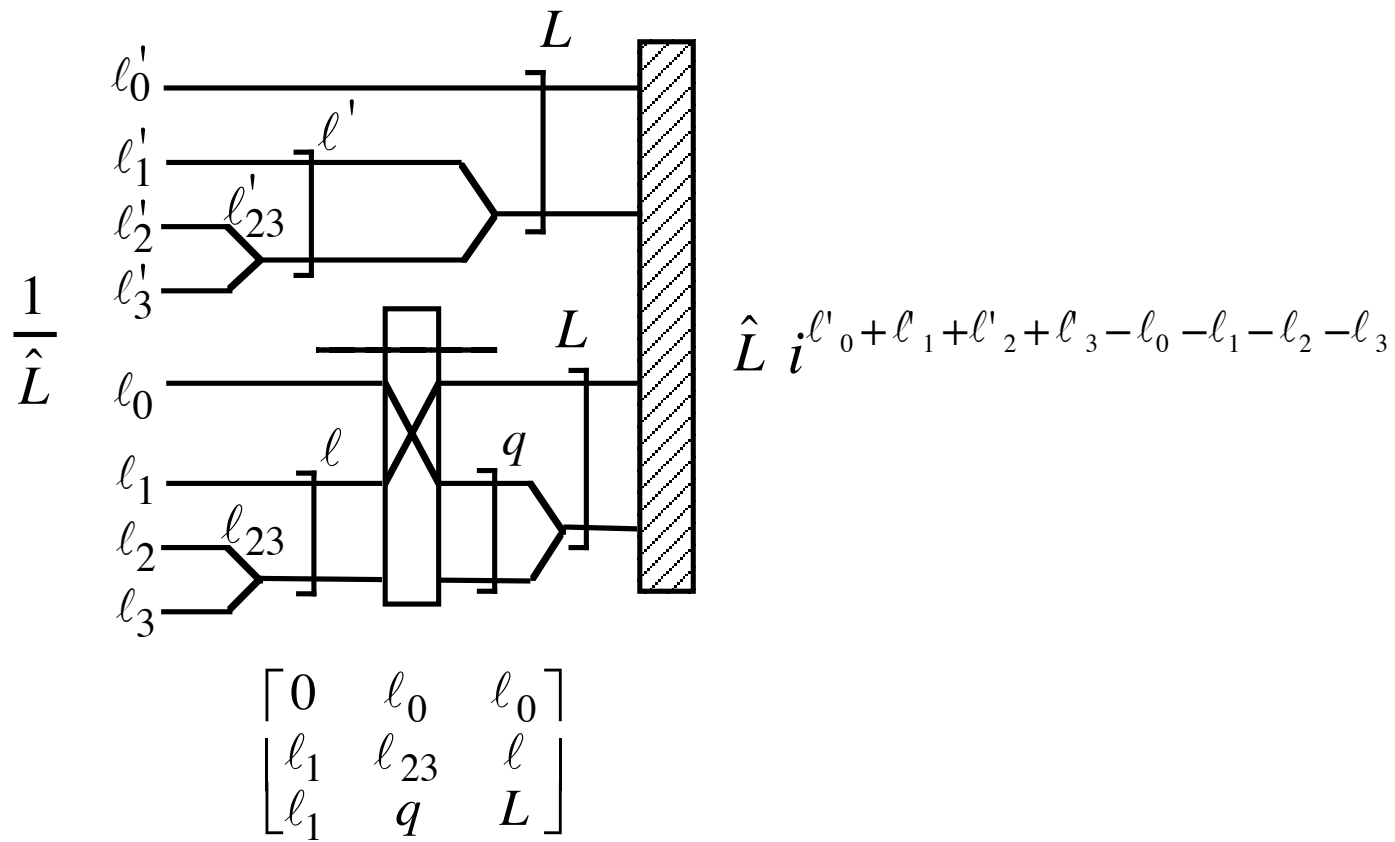

$$\frac{1}{\hat{L}} \qquad \hat{L}\, i^{\ell'_0 + \ell'_1 + \ell'_2 + \ell'_3 - \ell_0 - \ell_1 - \ell_2 - \ell_3}$$

$$\begin{bmatrix} 0 & \ell_0 & \ell_0 \\ \ell_1 & \ell_{23} & \ell \\ \ell_1 & q & L \end{bmatrix}$$

Figure 15